\documentclass[journal]{IEEEtran}
\usepackage{cite}
\usepackage{amsmath,amssymb,amsfonts}
\usepackage{algorithmic}
\usepackage{graphicx}
\usepackage{textcomp}
\usepackage{xcolor}
\usepackage{booktabs}
\usepackage{algorithm}
\usepackage{algorithmic}
\usepackage{caption}
\usepackage{subcaption}
\usepackage{subeqnarray}
\usepackage{cases}
\usepackage{amsmath}
\usepackage{dsfont}
\ifCLASSINFOpdf
\else
\fi

\begin{document}
\title{Learning Based Task Offloading in Digital Twin Empowered Internet of Vehicles}

\author{Jinkai~Zheng,~\IEEEmembership{Student Member,~IEEE,}%
	~Tom~H.~Luan,~\IEEEmembership{Senier Member,~IEEE,}%
        ~Longxiang~Gao,~\IEEEmembership{Senier Member,~IEEE,}%
        ~Yao~Zhang,%
        ~and~Yuan~Wu,~\IEEEmembership{Senier Member,~IEEE}%
\thanks{Jinkai~Zheng and Tom H. Luan are with the School of Cyber
	Engineering, Xidian University, Xian, Shaanxi, 710126 China (email:
	jkzheng@stu.xidian.edu.cn; tom.luan@xidian.edu.cn).}%
\thanks{Longxiang~Gao is with the School of Info Technology, Deakin University, Melbourne, Australia (e-mail:longxiang.gao@deakin.edu.au).}%
\thanks{Yao~Zhang is with the School of Telecommunications Engineering, Xidian University, Xian, Shaanxi, 710126 China (e-mail: yzhang\_01@stu.xidian.edu.cn).}%
\thanks{Yuan~Wu is with the Faculty of Science and Technology, University of Macau, Macau, China. (e-mail:yuanwu@um.edu.mo).}%
\thanks{Corresponding author: Tom~H.~Luan.}
}

%

\maketitle

\begin{abstract}	
Mobile edge computing has become an effective and fundamental paradigm for futuristic autonomous vehicles to offload computing tasks. However, due to the high mobility of vehicles, the dynamics of the wireless conditions, and the uncertainty of the arrival computing tasks, it is difficult for a single vehicle to determine the optimal offloading strategy. In this paper, we propose a Digital Twin (DT) empowered task offloading framework for Internet of Vehicles. As a software agent residing in the cloud, a DT can obtain both global network information by using communications among DTs, and historical information of a vehicle by using the communications within the twin. The global network information and historical vehicular information can significantly facilitate the offloading. In specific, to preserve the precious computing resource at different levels for most appropriate computing tasks, we integrate a learning scheme based on the prediction of futuristic computing tasks in DT. Accordingly, we model the offloading scheduling process as a Markov Decision Process (MDP) to minimize the long-term cost in terms of a trade off between task latency, energy consumption, and renting cost of clouds. Simulation results demonstrate that our algorithm can effectively find the optimal offloading strategy, as well as achieve the fast convergence speed and high performance, compared with other existing approaches.
\end{abstract}


\IEEEpeerreviewmaketitle

\section{Introduction}

\IEEEPARstart{T}{he Internet} of Vehicles (IoV) has become an important part of the intelligent transportation system by supporting in-vehicle applications, such as autonomous driving, augmented reality and infotainment services~\cite{IoV,AR,wang2018icast}. These applications are usually delay-sensitive and need to be completed under stringent delay constraints. Although the smart vehicles are equipped with computing units, the complex processing requirements of transportation applications may still pose a huge challenge to the vehicles.

Mobile edge computing (MEC) has been proposed to address problem between the limited computing resources of vehicles and the smooth completion of tasks for IoV and it uses its geographical advantages to provide more convenient services for vehicles. Therefore, a feasible solution is to perform task offloading in a heterogeneous computing scenario composed of vehicles, MEC server and cloud ~\cite{dai2018joint,wu2019task} to reduce latency or energy consumption~\cite{hu2018multi,hu2021mobility,nguyen2020modeling}. However, it is difficult for vehicles in the physical world to obtain global information about roads and other vehicles, and it is difficult for mobile vehicles to accurately obtain complex channel information and server resource state. In particular, the single vehicle cannot predict the arrival of future tasks due to the lack of global information. In this case, it is impossible to reserve computing resources in advance for a large number of tasks that may be received in the future, which brings great challenges to task offloading. A fundamental research issue is how to make a task offloading decision taking into account the arrival of future tasks to minimize the cost (e.g., execution time or energy consumption) of the system.
\begin{figure}[t]
	\centerline{\includegraphics[scale=0.25]{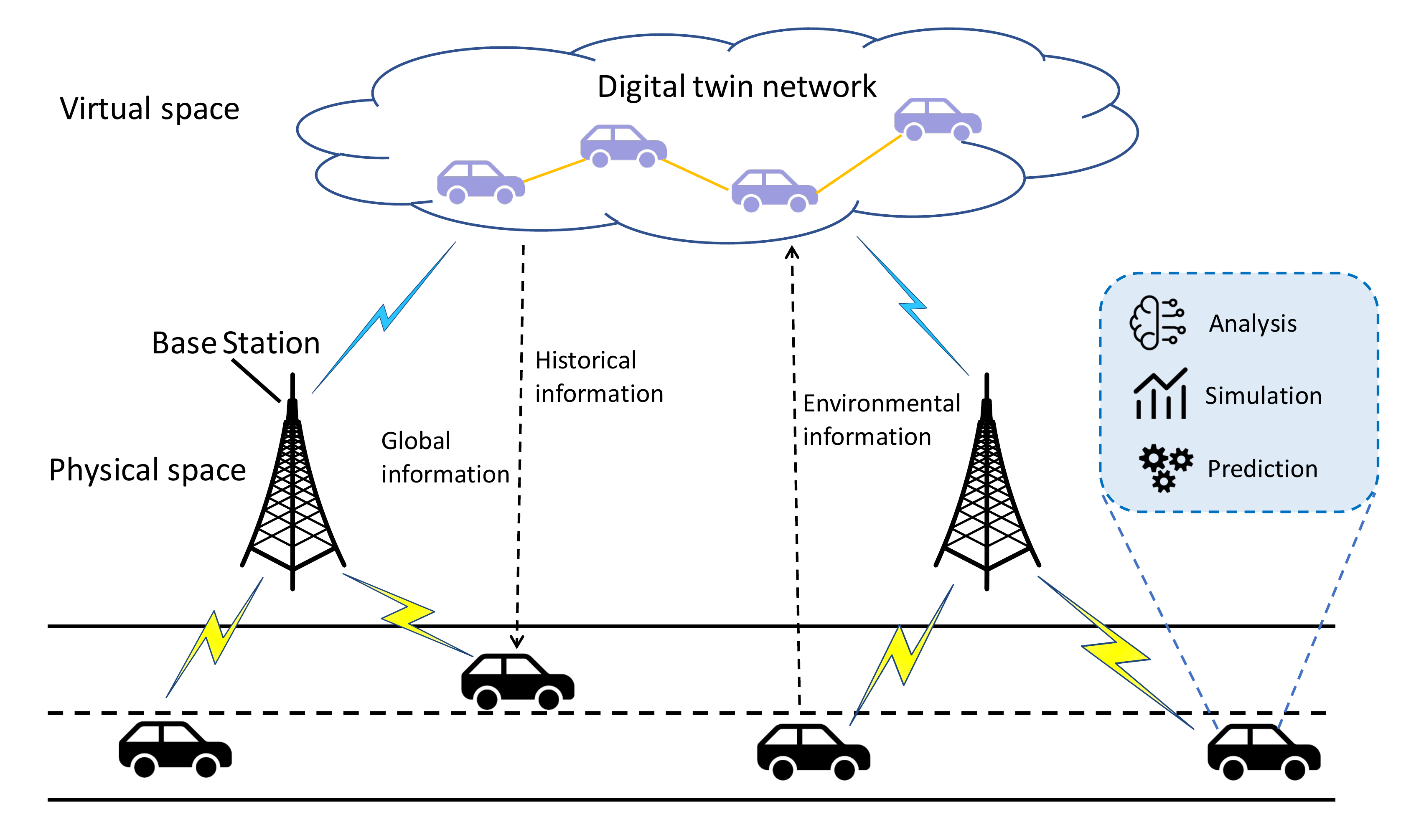}}
	\caption{Digital Twin empowered IoV framework.}
	\label{fig:DT}
\end{figure}

Digital twin (DT) emerges as an appealing technology to create digital replicas of physical objects~\cite{DigitalTwininIndustry,NASA,DT3}. The digital replicas (a.k.a., virtual models) are the mapping from the real world to the virtual space and constantly update parameters with real-time data from sensors. The virtual model can obtain the state of physical entities by sensing real-time data, and further provide users with accurate feedback by predicting, estimating and analyzing its dynamic changes~\cite{DT_pred}. 
With the support of DT technology, the virtual twin is generated and mapped to the physical vehicle through the Internet of Things, as shown in Fig.~\ref{fig:DT}. The vehicle in physical space is connected to the digital twin via the Internet. The DT empowered IoV collects state information of vehicles and surrounding environment through smart sensor devices~\cite{DT_share_info} and periodically synchronize the collected information to the digital twin through the network, so that the parameters of the virtual model are consistent with the physical vehicle~\cite{DT_vehicle1,DT_vehicle2}. In addition, information can be shared between digital twin networks in the cloud, and each vehicle can obtain information about the global network and road environment through through inter-twin communications. Accordingly, the global information can be used to simulate the road environment, analyze the current state and predict the future information to assist driving.

Therefore, task offloading with the assistance of DT is considered an effective solution in our work. This is because, as shown in Fig.~\ref{fig:DT}, (1) DT can obtain real-time global information through inter-twin communications~\cite{luan2021paradigm} including the driving path of the car, channel information and the status of heterogeneous computing units (i.e., MEC server or cloud). (2) DT can utilize historical information to predict the state changes of the system in a period of time in the future, and provide decision-making reference for driving to avoid possible extreme events (e.g., the arrival of unexpected tasks makes the task queue overflow due to the inability to process tasks in time). However, in the DT empowered IoV task offloading scenario, the traditional optimization method is difficult to deal with the complex system state. Specifically, first, when the vehicle is driving on the road, its position moves at all times. Secondly, the computing power of distributed computing units is constantly changing due to the occupancy of other vehicles. Thirdly, the wireless channel changes dynamically over time. These factors cause the state space to be huge, which is a challenge to traditional methods. Deep reinforcement learning (DRL) is an emerging technology to solve time-varying feature problems. In wireless networks and vehicle networks, DRL has shown excellent performance in optimizing computational offloading problems. Therefore, with the support of digital twins, we use DRL to find the best strategy for task offloading. 
\begin{figure*}[t]
	\centerline{\includegraphics[scale=0.21]{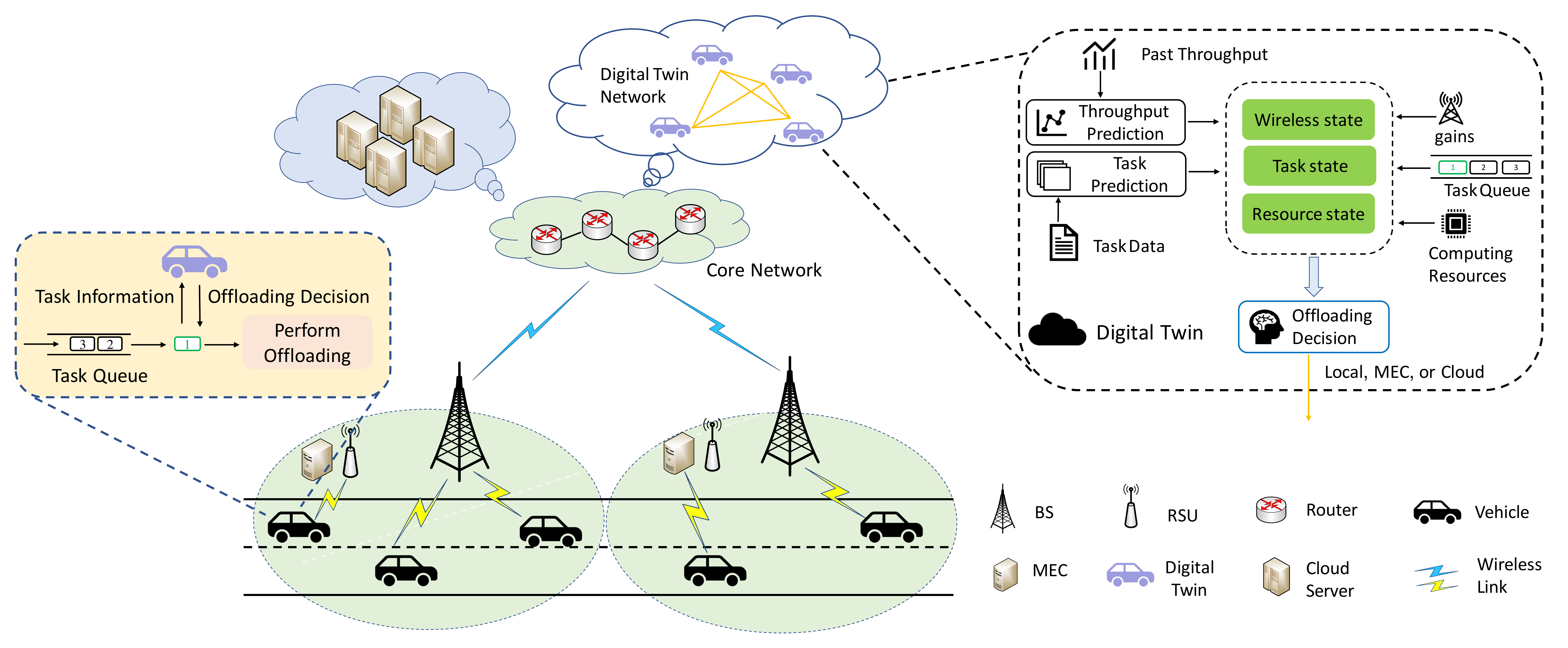}}
	\caption{Architecture of DT empowered task offloading for IoV.}
	\label{fig:DT_offloading}
\end{figure*}

In this paper, we investigate the task offloading problem in DT empowered IoV and resort to DRL technology to find the best offloading solutions. Different from the previous studies, we combine digital twins and DRL to capture and analyze the system states. The rental cost of cloud servers and bandwidth, dynamic wireless conditions, and the volatility of available computing resources in the MEC server are considered in this paper. Furthermore, we design a task prediction module to predict the arrival of future tasks to reserve valuable computing resources in advance and solve the problem of task queue overflow. In our work, we focus on minimize the weighted sum of latency, energy consumption, and renting of the offloading system. To address the problem, we model the task offloading process as a Markov Decision Process (MDP), which is an effective mathematical tool for modeling the behavior of a vehicle in a dynamic environment. Accordingly, each state in MDP is captured by the digital twin and used as input to DRL while the latency, energy consumption, and renting of the task are considered as the reward. Therefore, the goal of task offloading problem is to minimized accumulated discounted reward of the MDP. The main contribution of this paper can be summarized as follows:
\begin{itemize}
	\item We investigate a task offloading framework powered by digital twins where tasks in vehicles can be offloaded to different computing units (i.e., local vehicles, MEC server, or cloud). To address the issue, we model the offloading process as an MDP to balance task delay, energy consumption and cloud rental costs to minimize long-term costs.
	\item With the support of digital twins, we predict the arrival of future tasks, which can reserve valuable computing resources in advance at different levels for most appropriate computing tasks. In addition, we also consider the dynamics of the wireless channel and computing units, and the mobility of vehicles.
	\item We develop a DRL-based framework to handle huge state spaces and exploit asynchronous advantage actor-critic (A3C) algorithm to accelerate neural network training. Simulation results show that our algorithm is superior to existing solutions in terms of convergence efficiency and performance. In particular, the performance of our algorithm is significantly better than the algorithm without the application of digital twin technology.
\end{itemize}

The rest of paper is organized as follows. Section \uppercase\expandafter{\romannumeral2} describes related works and positions the original contributions of this paper. The system model is presented in Section \uppercase\expandafter{\romannumeral3}. The DRL based task offloading algorithm is proposed in Section \uppercase\expandafter{\romannumeral4}. Section \uppercase\expandafter{\romannumeral5} presents the evaluation method and analyses the performance of the algorithm. Finally, Section \uppercase\expandafter{\romannumeral6} closes this paper with conclusions.

\section{Realted Works}
Vehicle edge computing can realize the sharing of computing resources at the edge of the vehicle networks, and it has become a  dominant paradigm to meet the computing needs of smart car applications. In recent years, many researchers have made great efforts in the field of VEC. In~\cite{9051801}, the authors propose a two-stage meta-learning method to minimize the cost of consuming edge computing resources. Alahmadi \emph{et al}.~\cite{9203156} investigate the joint minimization of power consumption, propagation delay, and queuing delay when processing tasks are allocated to the network in the cloud-fog-VEC architecture. Lee \emph{et al}.~\cite{9097892} formulate the problem of allocating limited fog resources to vehicle applications in order to minimize the service delay and proposes a heuristic algorithm to effectively find the solution of the problem. In~\cite{9488886}, the authors investigate a service scenario of data-driven task offloading in a MEC-based vehicle network and designed an asynchronous deep q-learning algorithm to determine the offloading decision. Zhan \emph{et al}.~\cite{9026935} study a computational offload scheduling problem in a typical VEC scenario and they minimized long-term costs by weighing task delay and energy consumption. In the article, DRL is used to deal with the huge state space. 

In the DT empowered IoV, researchers have proposed various solutions to improve the QoS and safety of vehicle driving. In~\cite{9399641}, the authors propose a socially aware vehicle edge caching mechanism empowered by digital twin, which dynamically coordinates the caching capabilities of roadside units and smart vehicles based on user preference similarity and service availability. In~\cite{9304643}, to prevent potential dangers, the authors propose a new sensor fusion method that combines camera images and digital twin knowledge from the cloud to draw and match the bounding box of the target vehicle. In terms of  task offloading, Zhang \emph{et al}.~\cite{9451579} combine digital twin technology and artificial intelligence into the design of the automotive edge computing network and propose a coordination graph-driven vehicle task offloading scheme to minimize offloading costs. In \cite{9268482}, a service offloading method based on deep reinforcement learning is proposed to optimize the quality of service (QoS). In~\cite{9440709}, the author propose a digital twin-assisted real-time traffic data prediction method suitable for IoV applications to optimize traffic resource scheduling and alleviate traffic congestion that may occur during peak hours. Wang \emph{et al}.~\cite{9128938} propose a digital twin framework for connected vehicles using vehicle-to-cloud (V2C) communication so that the driver can control the vehicle in a smarter way. 

The above research provide a variety of useful and promising solutions for task offloading under the Internet of Vehicles. However, some of the existing solutions are based on the assumption of a static environment, and some do not consider the cost of renting cloud services. More importantly, these methods do not take into account the arrival of future tasks. In this paper, we propose an offloading framework based on a dynamic environment, which fully considers the dynamics of computing units and wireless communications, and possible rental costs. Especially we consider carefully the arrival of future tasks. 
\section{System Model}
In this section, we describe the service architecture for task offloading in DT empowered IoV and model three different types of transmission and computation for computing vehicles in detail.
\subsection{System Description}
We consider a scenario, as shown in Fig.~\ref{fig:DT_offloading}. Vehicles are covered by RSUs and base stations (BSs), and each RSU is equipped with MEC server to process the tasks received from vehicles. We assume that vehicles can communicate with RSUs through wireless connection, and the coverage of each RSU does not overlap. In addition, vehicles can communicate with cloud servers and the digital twin through BSs along the road. The digital twin of each vehicle has an offloading decision making module, and the twins can exchange data to obtain global information. For each vehicle, a digital twin of it self is generated with information of position, speed, vehicle gap, and dashcam videos collected by vehicular sensors and cameras. The execution method (offloading decision) of each task in the vehicle is determined with the assistance of the digital twin in cloud\footnote{Making the offloading decision by a digital twin with powerful computing power instead of local vehicle can save more time and energy.}. Specifically, for a certain task, the agent in digital twin makes the offloading decision by observing the current state and sends the decision to local vehicle for execution. The current state includes wireless state, tasks state and resources state. The wireless state consists of channel gains and predicted throughput while tasks state consists of the task information (i.e., data size and computing resources required by the task) and predicted task arrival state. For resources state, it includes the computing resource information of MEC server and the energy information of the vehicle. When this task is executed, the agent in digital twin makes an offloading decision for the next task based on the current state, and updates the parameters with vehicle in time. According to~\cite{9026935}, a vehicle can be only served by one RSU, so when the vehicle drives out of the RSU's communication coverage, a handover occurs. Therefore, in order to improve channel utilization, we assume that task data (except for execution results) are not allowed to be transmitted between RSUs (i.e., when the vehicle drives out of the current RSU's communication coverage, tasks that have not been transmitted will be discarded).  

In the DT empowered IoV scenario, vehicular terminals are denoted by a set $\mathcal{V}=\left\{ v_{1},v_{2},...,v_{N}\right\} $, and MEC servers are denoted by $\mathcal{M}=\left\{ m_{1},m_{2},...,m_{K}\right\} $. Let $\left\Vert Q\right\Vert $ denotes the task queue length and $q(t)$ is the occupied size of the task queue at time slot $t$. The average distance between adjacent RSUs is assumed to be $L$ meters. We assume that the parameters of the digital twin keep pace with the vehicle, and the digital twin uses a 5-tuple to characterize the vehicle $v$ in the physical world at time slot $t$, which can be expressed as 
\begin{equation}
DT_{v}\left( t\right) =\left\{ Q_{v}\left( t\right)
,f_{v}^{l},loc_{v}^{l}\left( t\right) ,\Gamma (t),\Phi _{v}\left( t\right)
\right\},
\end{equation}
where $Q_{v}\left( t\right) $ is the queue state, $f_{v}^{l}$ denotes computing resources of local vehicle, $loc_{v}^{l}\left( t\right) $ is the current location of $v$, $\Gamma (t)$ is the task parameters at time $t$  (e.g., the data size and required computing resources of the tasks in the task queue), and $\Phi _{v}\left( t\right) $ denotes other parameters related to the vehicle.

For each task $i$ in task queue, it can be represented as $\left\{ d_{i},cr_{i},T_{i}^{\max }\right\} $. Here, $d_{i}$ represents the input data size of this task. $cr_{i}$ (in CPU cycles) denotes the amount of computing resources required to execute task $i$. $T_{i}^{\max }$ is the maximum delay tolerance for the task. We assume that the system adopts first-come-first-service scheduling method and the task being executed cannot be preempted by other tasks. At the beginning of each scheduling, each digital twin makes its own decisions from
$\left\{ 0,1,2\right\}$ for task $i$ based on current state information where $\left\{ 0\right\} $ to $\left\{ 2\right\} $ represent local execution, MEC execution, cloud execution, respectively.

In terms of task queue, we model the arrival of the tasks as a Poisson process with rate $\lambda$. In practice, the complex road environment (i.e., a large number of vehicles and buildings) or the driver's frequent driving command switching (i.e., braking, steering, lane changing) will cause the number of tasks to surge in a short period of time, corresponding to a higher $\lambda$.

\subsection{Channel Model}
In this section, we formulate the channel model of MEC servers, cloud and vehicles. Different from most of the previous work, our work is based on dynamic channel environment, and the past channel gains are important reference to make offloading decisions.

We consider a complete synchronous time slot system with duration $T$, and each vehicle is equipped with a single antenna. We refer to~\cite{blockfading} to use the block fading model to express the channel gains, and assume that the channel gains remain unchanged for the same duration. Accordingly, the channel gain between $v_{n}$ and MEC server $m_{k}$ at time slot $t$ is denoted by
\begin{equation}
g_{n\rightarrow k}^{t}=\left\vert h_{n\rightarrow k}^{t}\right\vert
^{2}\vartheta _{n\rightarrow k}^{t}, \ t=1,2,...,\label{equ:channel_gain}
\end{equation}
where $h_{n\rightarrow k}^{t}$ denote the small-scale fading~\cite{smallscalefading}, and $\vartheta _{n\rightarrow k}^{t}$ is the large-scale path loss and shadowing. We adopt a block fading channel model which is general to our  environment (i.e., Jakes model), and use the first-order Gauss-Markov process to describe the small-scale fading component. The update of $h_{n\rightarrow k}^{t}$ is given as
\begin{equation}
h_{n\rightarrow k}^{t}=\kappa h_{n\rightarrow k}^{t-1}+l_{n\rightarrow
	k}^{t}\label{equ:small_scale},
\end{equation}
where $h_{n\rightarrow k}^{0}\sim CN\left( 0,1\right)$ is circularly symmetric complex Gaussian distribution unit variance, and $l_{n\rightarrow k}^{1},l_{n\rightarrow k}^{2},...$ denote the channel innovation process with
distribution $\mathcal{CN}\left( 0,1-\kappa ^{2}\right) $. According to the the Jakes model, the correlation $\kappa$ can be defined by
\begin{equation}
	 \kappa =J_{0}\left( 2\pi f_{D}T\right) \label{equ:correlation},
\end{equation}
where $J_{0}\left( \cdot \right) $ is the zeroth order Bessel function. The maximum Doppler frequency is defined as $f_{D}=vf_{c}/c$, where $v$ is the speed of vehicle, $f_{c}$ being the carrier frequency, $c=3\times 10^{8}m/s$ is transmission rate of electromagnetic waves in the air, and $T$ denotes the CSI feedback interval.

The wireless communication data rate between $v_{n}$ and the currently connected $m_{k}$ can be expressed as
\begin{equation}
C_{n,k}^{t}=B_{n,k}\log _{2}\left( 1+\frac{g_{n\rightarrow k}^{t}p_{n}^{\textrm{tr}}
}{\sigma ^{2}}\right) \label{equ:mec_rate}.
\end{equation}
Here, $\sigma^{2}$ is the additive white Gaussian noise power spectral density, $B_{n,k}$ denotes the available bandwidth allocated by $m_{k}$ to $v_{n}$, and $p_{n}^{\textrm{tr}}$ is transmission power of vehicle.

Follow the same solution, we can express the achievable transmission rate between vehicle $n$ and cloud as
\begin{equation}
C_{n,c}^{t}=B_{n,c}\log _{2}\left( 1+\frac{g_{n\rightarrow c}^{t}p_{n}^{\textrm{tr}}
}{\sigma ^{2}}\right) \label{equ:cloud_rate},
\end{equation}
where $g_{n\rightarrow c}^{t}$ denotes the channel gain between vehicle $n$ and cloud server, and $B_{n,c}$ is bandwidth that cloud allocate to the vehicle.

\subsection{Local Computing Model}
In this section, the task $i$ will be performed on the local vehicle, and the offloading action received by vehicle from the digital twin is $a_{i}=0$. In this case, the total cost of this task includes only the execution time and energy consumption of this task by vehicular computing units. The execution time is determined by the computing power of the vehicle and the computing resources required by the task, which has nothing to do with the size of the input data of the task. For simplicity, we assume that the computing resource of the vehicle is a constant, and the execution time of task $i$ in vehicle $n$ can be expressed as
\begin{equation}
	T_{n,i}^{\textrm{com}}=\frac{cr_{i}}{f_{n}^{\textrm{l}}}\label{equ:local_com_time},
\end{equation}
where $cr_{i}$ denotes the computing resources required by $i$, and $f_{n}^{\textrm{l}}$ (i.e., CPU cycles per time slot) is computing resources of vehicle. In~(\ref{equ:local_com_time}), $f_{n}^{\textrm{l}}$ satisfies $p_{n}=\zeta \left( f_{n}^{\textrm{l}}\right) ^{\tau }$, where $\zeta$ is the effective switched capacitance depending
on the chip architecture, and $\tau$ is a constant. Therefore, given the energy consumption of unit time slot, denoted by $p_{n}$, local energy consumption for computing task $i$ is given by
\begin{equation}
	E_{n,i}^{\textrm{com}}=p_{n}T_{n,i}^{\textrm{com}}.
\end{equation}
\subsection{MEC Computing Model}
In this section, the offloading decision made by the digital twin based on the current environmental state is $a_{i}=1$, which means that the task will be transmitted to the MEC server for execution. Different from local computing model, the total cost is composed of the cost of task transmission and the cost of task calculation. Moreover, the model also considers the dynamics of wireless communication and the availability of MEC server computing resources. In this model, if the digital twin of vehicle $n$ decide to offload task $i$ to MEC $k$ at time slot $t$, the transmission time can intuitively be expressed as
\begin{equation}
T_{n,i,k}^{\textrm{tr}}=\frac{d_{i}}{C_{n,k}^{t}}\label{equ:err_trans_time},
\end{equation}
where $d_{i}$ is input data of task $i$. Note that a task may require multiple time slots to complete the transmission from the vehicle to the MEC server. If the transmission is not completed within time slot $t$, the transmission rate will change at time slot $t+1$. Therefore,  (\ref{equ:err_trans_time}) cannot meet the calculation requirements. Mathematically, the complete transmission process of task $i$ from vehicle $n$ to MEC $k$ should satisfy 
\begin{equation}
\int\nolimits_{t_{\textrm{start}}}^{t_{\textrm{end}}}C_{n,k}^{t}dt=d_{i}.
\end{equation}
Here, the transmission time can be obtained by $t_{\textrm{end}}-t_{\textrm{start}}$. In our work, we use an iterative method to calculate it. Specifically, we assume that the vehicle decides to offload task $i$ to an MEC server at time slot $t_{\textrm{start}}$, and we initialize indicator $D_{i}^{t_{\textrm{start}}}=d_{i}$. Accordingly, the update of $D_{i}^{t}$ follows
\begin{equation}
D_{i}^{t}=\max \left\{ D_{i}^{t-1}-T\cdot C_{n,k}^{t-1},0\right\} \label{equ:reamain_data_process},
\end{equation}
where $D_{i}^{t}$ denotes the remaining input data size of $i$ at time slot $t$, and $T$ is the time interval between two adjacent time slots. We assume at time slot $t_{\textrm{end}}$ that $D_{i}^{t_{\textrm{end}}}=0$ $\left( t_{\textrm{end}}\geqslant t_{\textrm{start}}\right) $, the transmission time of task $i$ can be defined as
\begin{equation}
T_{n,i,k}^{\textrm{tr}}=t_{\textrm{end}}-t_{\textrm{start}}.
\end{equation}

When the vehicle transmits tasks to the MEC server, it consumes energy. In practice, vehicles are energy-constrained devices and energy consumption is a factor that must be considered. The energy consumption of transmitting task $i$ from vehicle $n$ to MEC $k$ is given by
\begin{equation}
	E_{n,i,k}^{\textrm{tr}}=p_{n}^{\textrm{tr}}T_{n,i,k}^{\textrm{tr}},
\end{equation}
where $p_{n}^{tr}$ is transmission power of vehicle $n$.

After being transmitted to the MEC server, the task will be executed immediately. Note that MEC server deployed on RSU usually serves multiple vehicles at the same time, which will cause the available computing resources of the MEC server to be different in different time slots. Therefore, we model the available computing resources of the MEC server as a stochastic process to analyze a single vehicle. Similar to (\ref{equ:reamain_data_process}), we have $F_{i}^{t_{\textrm{start}}}=cr_{i}$, and the update of $F_{i}^{t}$ follows
\begin{equation}
F_{i}^{t}=\max \left\{ F_{i}^{t-1}-T\cdot f_{k}^{t-1},0\right\},
\end{equation}
where $F_{i}^{t}$ and $f_{k}^{t}$ denote the remaining computation resource required to complete task $i$ and the computing resources of MEC server $k$ at time slot $t$, respectively. At $t_{\textrm{end}}$, we assume that $F_{i}^{t_{\textrm{end}}}=0$ $\left( t_{\textrm{end}}\geqslant t_{\textrm{start}}\right) $, and  execution time of completing task $i$ on MEC server $k$ is expressed as follows
\begin{equation}
	T_{n,i,k}^{\textrm{com}}=t_{\textrm{end}}-t_{\textrm{start}}.
\end{equation}
In MEC computing model, we ignore the energy consumption of task execution because it has nothing to do with the cost of the vehicle. 

\subsection{Cloud Computing Model}
In this model, for task $i$, the digital twin sends the offloading action $a_{i}=2$ to the vehicle, and the vehicle transmits task $i$ to the cloud server via the Internet through the currently connected base station (BS). The transmission time of task $i$ from vehicle $n$ to cloud can be written as
\begin{equation}
	T_{n,i,c}^{\textrm{tr}}=\frac{d_{i}}{C_{n,c}^{t}}\label{equ:cloud_trs}.
\end{equation}
Based on (\ref{equ:cloud_trs}), the energy consumption of task transmission is expressed as follows.
\begin{equation}
	E_{n,i,c}^{\textrm{tr}}=p_{n}^{\textrm{tr}}T_{n,i,c}^{\textrm{tr}}.
\end{equation}
Further, the cost of using Internet can not be neglected. We assume the unit
cost of using Internet (i.e., dollar per bit) is $\varpi _{tr}$, then the cost of transmitting task $i$ can be expressed as
\begin{equation}
	rc_{n,i,c}^{\textrm{tr}}=\varpi _{tr}\cdot d_{i}.
\end{equation}
Usually, the cloud server is composed of many servers with powerful computing resources, so we assume that the cloud computing resources are constant (denoted by $f_{c}$) during the execution of the task. The computation time of completing task $i$ on cloud is given as
\begin{equation}
	T_{n,i,c}^{\textrm{com}}=\frac{cr_{i}}{f_{c}}.
\end{equation}
Further, let $\varpi _{com}$ denote the cost of using clouds, which is related to the computing resources required by tasks. The computation cost of task $i$ can be expressed as
\begin{equation}
	rc_{n,i,c}^{\textrm{com}}=\varpi _{\textrm{com}}\cdot \left( cr_{i}\right) ^{\mu },
\end{equation}
where $\mu \geqslant 1$ is the price factor related to the cloud service provider.

\subsection{Objective}
No matter which of the three actions the digital twin takes, for a task, the total cost includes execution time, energy consumption, and rental cost (if the task is offloaded to the cloud). Therefore, for tasks $i$, we model the time cost as 
\begin{equation}
	t_{i}=\left\{ 
	\begin{array}{c}
		T_{n,i}^{\textrm{com}}, \\ 
		T_{n,i,k}^{\textrm{tr}}+T_{n,i,k}^{\textrm{com}}, \\ 
		T_{n,i,c}^{\textrm{tr}}+T_{n,i,c}^{\textrm{com}},
	\end{array}
	\right. 
	\begin{array}{c}
		if \quad a_{i}=0 \ , \\ 
		if \quad a_{i}=1  \ , \\ 
		if \quad a_{i}=2 \ ,
	\end{array}
\label{equ:t_i}
\end{equation}
where $n$ and $k$ are objective vehicle and MEC server, respectively. Similarly, the energy cost and rental cost are formulated as follows, respectively.
\begin{equation}
	e_{i}=\left\{ 
	\begin{array}{c}
		E_{n,i}^{\textrm{com}}, \quad \quad \quad \quad  \\
		E_{n,i,k}^{\textrm{tr}}, \quad \quad \quad \quad \\ 
		E_{n,i,c}^{\textrm{tr}}, \quad \quad \quad \quad
	\end{array}
	\right. 
	\begin{array}{c}
		if \quad a_{i}=0  \ , \\ 
		if \quad a_{i}=1  \ , \\ 
		if \quad a_{i}=2  \ ,
	\end{array}
\label{equ:e_i}
\end{equation}
\begin{equation}
	rc_{i}=\left\{ 
	\begin{array}{c}
		0, \\ 
		0, \\ 
		rc_{n,i,c}^{\textrm{tr}}+rc_{n,i,c}^{\textrm{com}},
	\end{array}
	\right. 
	\begin{array}{c}
		if \quad a_{i}=0 \ , \ \\ 
		if \quad a_{i}=1  \ , \\ 
		if \quad a_{i}=2 \ , 
	\end{array}%
\label{equ:rc_i}
\end{equation}
From (\ref{equ:t_i}), (\ref{equ:e_i}), and (\ref{equ:rc_i}), we can express the total cost for task $i$ in a vehicle as 
\begin{equation}
	C_{i}=\xi _{1}t_{i}+\xi _{2}e_{i}+\xi _{3}rc_{i} \label{equ:total_cost}.
\end{equation}
Here, $\xi _{1}$, $\xi _{2}$, and $\xi _{3}$ are weighting factors between task execution latency, energy consumption, and the rental cost respectively, which represents the importance of each cost to the total cost of the task. 

In digital twin empowered task offloading system, the cost of competing a task is evaluated from execution time, energy consumption, and rental cost. Our goal is to minimize the long-term cost of the tasks. For simplicity, we model the computing resources of MEC server as a stochastic process, which represents the state of computing resources of MEC server being occupied by other vehicles at time slot $t$ (i.e., available computing resources at time slot $t$). Therefore, our optimization problem is transformed into minimizing the long-term average cost of  tasks in a single vehicle, which can be defined as 
\begin{subequations}
	\label{objective}
	\begin{align}
		\min\limits_{a_{i}\in \left\{ 0,1,2\right\} }&\lim\limits_{w\rightarrow
			\infty }\frac{1}{w}\sum\limits_{i=1}^{w}C_{i}\nonumber \\
		&=\min\limits_{a_{i}\in \left\{ 0,1,2\right\} }\lim\limits_{w\rightarrow
			\infty }\frac{1}{w}\sum\limits_{i=1}^{w}\xi _{1}t_{i}+\xi _{2}e_{i}+\xi
		_{3}rc_{i},\label{equ:obj}\\
		s.t. \quad& \sum\limits_{j=1}^{3}\xi _{j}=1 \label{c:1}, \\
		&a_{i}\in \left\{ 0,1,2\right\} ,\forall i\in Z^{+}, \label{c:2} \\
		&t_{i}\leqslant T_{i}^{\max },\forall i\in Z^{+}, \label{c:3} \\
		&0\leqslant f_{k}^{t}\leqslant f_{k}^{\max },\forall k\in M,t=1,2,...,\label{c:5} \\
		&0\leqslant f_{c}\leqslant f_{c}^{\max },\label{c:6}\\
		&0\leqslant p_{n}^{\textrm{tr}}\leqslant p_{n}^{\textrm{tr},\max },\label{c:7}\\
		&q\left( t\right) \leqslant \left\Vert Q\right\Vert , t=1,2,..., \label{c:8}\\
		&rp(t)>0,t=1,2,... \label{c:9}
	\end{align}
\end{subequations}
where constraint (\ref{c:1}) shows that the sum of the weight coefficients is $1$. (\ref{c:2}) indicates that the action can only be taken from local vehicle, MEC server, and cloud. (\ref{c:3}) ensures that the execution time of each task cannot be greater than the time that the system can tolerate. (\ref{c:5}) and (\ref{c:6}) are the maximum computing resources constraints of MEC server and cloud allowed to vehicles. (\ref{c:7}) indicates the constraint of the maximum transmission power of vehicles. (\ref{c:8}) shows that at time slot $t$, the data size of tasks in vehicle cannot exceed the queue length. (\ref{c:9}) indicates that the remaining power must be greater than zero at time slot $t$. Probelm (\ref{equ:obj}) is a mixed integer non-linear problem where it is difficult to be solved in polynomial time. Fortunately, DRL can solve this kind of problem effectively. Hence, we adopt it to solve the optimization problem in (\ref{equ:obj}) in the following section.

\section{Proposed DRL based offloading algorithm}
\subsection{DRL Background}
Reinforcement learning (RL) refers to a machine learning paradigm in which RL agents learn the optimal policy through trial and error with the environment for sequential decision-making problems. The environment is usually modeled as an MDP, which can be defined as
\begin{equation}
	\left\langle \mathcal{S},\mathcal{A},\mathcal{R},\mathcal{P},\gamma \right\rangle,
\end{equation}
where $\mathcal{S}$ is a finite set of states, $\mathcal{A}$ is a finite set of actions, $\mathcal{R}$ is reward function, $\mathcal{P}$ denotes a state transition probability matrix, and $\gamma \in \left[ 0,1\right] $ is a discount factor.

Combining reinforcement learning with deep learning, i.e., DRL, enables the agent to process complex state spaces and obtain better performance. In RL problem, the agent will continue to interact with the environment. At each step, the agent observes current state $s_{i}$ from the state space $\mathcal{S}$, and take an action $a_{i}$ in action space $\mathcal{A}$ under the policy $\pi$, which is a specific mapping from the state $s_{i}$ to the action $a_{i}$. After this, the environment (i.e., MDP) transits to the next state $s_{i+1}$ and agent obtain the reward $r_{i}$. The state transition process can be denoted as $P\left( s_{i+1}\mid s_{i},a_{i}\right) $ and the reward function is $R\left( s_{i+1},s_{i},a_{i}\right) $. The goal of agent is to maximize the accumulated reward from state $s_{i}$, which is defined as
\begin{equation}
	G_{i}=r_{i}+\gamma r_{i+1}+\ldots =\sum\limits_{l=0}^{\infty }\gamma
	^{l}r_{i+l}.
\end{equation}

Note that in our scenario, the state transition of the system does not depend on the model (i.e., $\mathcal{P}$ is unknown), and we call it model free method. Therefore, we will focus on model-free reinforcement learning. Accordingly, the RL algorithms can be divided into value-based methods and policy-based methods. 

For value-based methods, the value function can be defined as $V_{\pi }\left( s\right) =E_{\pi }\left[ G_{i}\mid s_{i}=s\right] $, which is the expected accumulated reward starting from state $s$ under policy $\pi$. The value function $V_{\pi }\left( s\right) $ represents how good the state $s$ is, which is part of our proposed algorithm (i.e., critic network). The action-value function $q_{\pi }\left( s,a\right) =E_{\pi }\left[ G_{i}\mid s_{i}=s,a_{i}=a\right] $ is the expected accumulated reward starting from state $s$ and taking action $a$ under policy $\pi$. Therefore, the relationship between $V_{\pi }\left( s\right) $ and $q_{\pi }\left( s,a\right) $ can be written as $V_{\pi }\left( s\right) =\sum\limits_{a\in A}\pi \left( s,a\right) q_{\pi
}\left( s,a\right) $. Based on these, RL algorithm needs to solve the optimal state-value function $	V_{\ast }\left( s\right) =\max\limits_{\pi }V_{\pi }\left( s\right)$ and the optimal action-value function $q_{\ast }\left( s,a\right) =\max\limits_{\pi }q_{\pi }\left( s,a\right) $ for any state. In addition, the value function can be  approximated by Deep Neural Network (DNN) to solve large problems. We assume that $\theta$ is the parameter of DNN, the value function can be approximated as $V\left( s,\theta \right) =V_{\pi }\left( s\right) $. Therefore, the goal of agent is to find best $\theta$ to minimize the value function. 

For policy-based methods, the policy $\pi \left( s,a\right) =P\left[ a\mid s\right] $ refers to the probability of taking action $a$ at state $s$ following policy $\pi$. This method is effective when dealing with high-dimensional spaces and can learn stochastic policies, which is another part of our proposed algorithm (i.e., actor network). Similarly, the policy can be approximated as $\pi _{\theta }\left( s,a\right) =P\left[ a\mid s,\theta \right] $. Given $\pi _{\theta }\left( s,a\right) $, the goal of agent is to find best $\theta$ to minimize the objective function. 

\begin{figure}[t]
	\centerline{\includegraphics[scale=0.4]{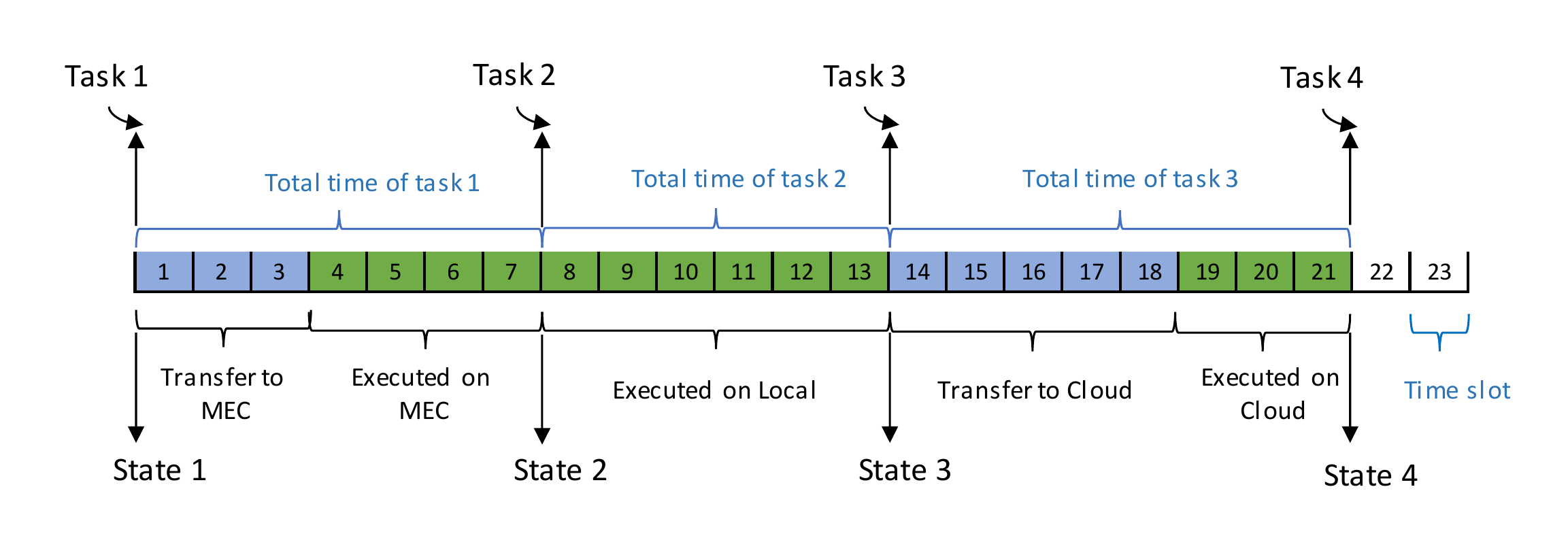}}
	\caption{An example of task offloading timeline. Time slots between any two states represent the total time to perform a task.}
	\label{fig:time_slots}
\end{figure}
\subsection{Algorithm Design and Network Architecture}
We discuss the details of the algorithm design. Firstly, we formulate the task offloading problem as an MDP. Secondly, we define the state, action and reward in MDP according to the offloading scenario. Finally, we describe in detail the structure of the neural network used to solve the optimal policy for MDP.

\begin{figure}[t]
	\centerline{\includegraphics[scale=0.4]{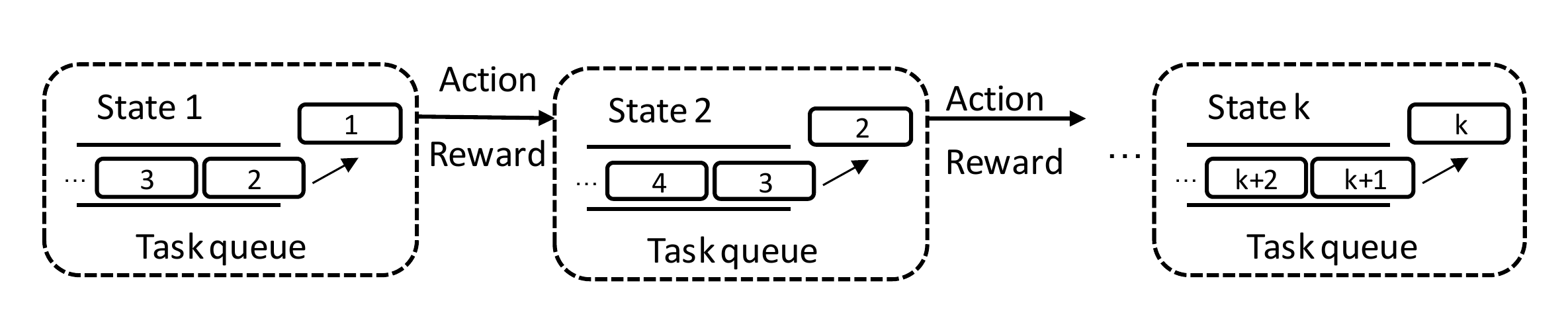}}
	\caption{Markov decision process model for task offloading.}
	\label{fig:MDP}
\end{figure}

\textbf{State. }
The offloading decision $a_{i}$ for task $i$ is made based on state $s_{i}$. According to existing solutions, each time slot can be considered as a state of the environment.  However, most tasks require multiple time slots to complete, i.e., each time slot in the task execution or transmission process is also considered as the state of the system. Fig.~\ref{fig:time_slots} shows an example of the timeline of task offloading. We can see that multiple time slots are required for task transmission and execution, corresponding to multiple states in MDP. These states are generated by the system and are not related to the actions taken by the agent. In other words, the offloading action only affects the state when this action is take, e.g., time slot $1$, $8$, $14$, and $22$. Accordingly, we only consider time slot when each action is taken (i.e., we remove the process of task execution and transmission from MDP) to simplify the state space, which can make training more effective. The MDP for task offloading is shown in Fig.~\ref{fig:MDP}. For task $i$ at state $s_{i}$, the agent takes an action $a_{i}$ after observing the current state. When the task is completed after one or more time slots, the MDP transfers to the next state $s_{i+1}$, and the agent gets the reward for this action. The state $s_{i}$ at time slot $t$ is given by
	\begin{align}
	s_{i}=&[ q\left( t\right) ,rp\left( t\right) ,d_{i},cr_{i},G_{m}\left(
	t\right) ,G_{c}\left( t\right) ,F_{m}\left( t\right) , \nonumber \\
	&T_{m}^{\textrm{pr}},T_{c}^{\textrm{pr}} ,U_{\textrm{task}}^{\textrm{pr}}],
\end{align}
where $q(t)$ denotes total size of pending tasks before scheduling (i.e., the occupied size of the task queue at time slot $t$), and $rp(t)$ is the remaining power of vehicle. $d_{i}$ and $cr_{i}$ are input data size and computing resources required by task $i$, respectively. $G_{m}\left( t\right) =\left[ g_{m}^{t-j},g_{m}^{t-j+1},...,g_{m}^{t}\right] $ are gains between vehicle and the connected MEC server in past $j$ time slots, and $G_{c}\left( t\right) =\left[ g_{c}^{t-j},g_{c}^{t-j+1},...,g_{c}^{t}\right] $ are gains between vehicle and cloud server in past $j$ time slots. With $G_{m}(t)$ and $G_{c}(t)$, the agent can analyze the changes in wireless channel over a period of time in the future. $F_{m}\left( t\right) =\left[ f_{m}^{t-u},f_{m}^{t-u+1},...,f_{m}^{t}\right] $ denotes the available computing resources of MEC server in past $u$ time slots, which can reflect the changes in the computing resources of MEC server in the past period of time. In other words, $F_{m}(t)$ represents the amount of tasks offloaded by other vehicles to the MEC server in the past $u$ time slots. $T_{m}^{\textrm{pr}}$ and $T_{c}^{\textrm{pr}}$ are predicted average throughput of the next $w$ time slots from the throughput prediction module, which is a recurrent neural network (RNN) discussed later. Specifically, $U_{\textrm{task}}^{\textrm{pr}}$ denotes the predicted task arrival state powered by RNN.

\textbf{Action.}
Digital twin can make three types of actions: local execution, MEC execution, and cloud execution. The offloading action $a_{i}$ for task $i$ based on state $s_{i}$ can be take from action space, defined as $\mathcal{A}=\left\{ 0,1,2\right\}$.

\textbf{Reward. }
In DT empowered task offloading system, our goal is to minimize the cost of tasks. Therefore, the reward should be a monotonically decreasing function of the cost, i.e.,  the lower execution time, energy consumption, and renting cost will bring higher reward. In addition, we define the penalties for execution time and task queue overflow. First, task execution overtime will bring bad Quality of Experience (QoE) and even threaten driving safety. Therefore, we define the penalty for execution time of task $i$ as
\begin{equation}
	P_{\textrm{t},i}=\max \left\{ t_{i}-T^{\max },0\right\}.
\end{equation}

Second, for a vehicle, the task queue capacity is limited. If the task arrival rate exceeds the scheduling rate, the queue will overflow due to the accumulation of tasks over time. When the queue overflows, new tasks are discarded, and the vehicle may have accidents due to lack of important data (e.g., vehicle control decisions). The penalty for queue overflow event is defined as
\begin{equation}
	P_{\textrm{over},i}=\max \left\{ \left[q\left( t\right) +\sum\limits_{j=t}^{t^{\prime
	}}n_{j}-d_{i}\right] -\left\Vert Q\right\Vert ,0\right\}.
\end{equation}
Here, $n_{j}$ is total data size of tasks received by vehicle at time slot $j$ and  $t^{\prime}-t$ denotes the number of time slots to complete task $i$ from time slot $t$ (i.e., task $i$ is scheduled at time $t$ and completed at $t^{\prime}$). 
Accordingly, to make the cumulative reward in line with our optimization objective, the  reward for task $i$ (state $i$ in MDP) is defined as
\begin{equation}
	r_{i}=\left\{ 
	\begin{array}{c}	
		-F,\\
		-\upsilon \left[ C_{i}+\eta P_{t,i}+\psi P_{\textrm{over},i}\right] ,
	\end{array}%
	\right. 
	\begin{array}{c}
		\textrm{if} \ \textrm{handover,}\\ 
		\textrm{else.}%
	\end{array}\label{equ:reward}
\end{equation}

where $F\in R^{+}$ denotes a penalty for handover event. $\eta$ and $\psi $ are the penalty coefficients for execution time and task queue overflow, respectively. $\upsilon \in R^{+}$ is used to scales the value range of the reward. 

\begin{figure}[t]
	\centerline{\includegraphics[scale=0.4]{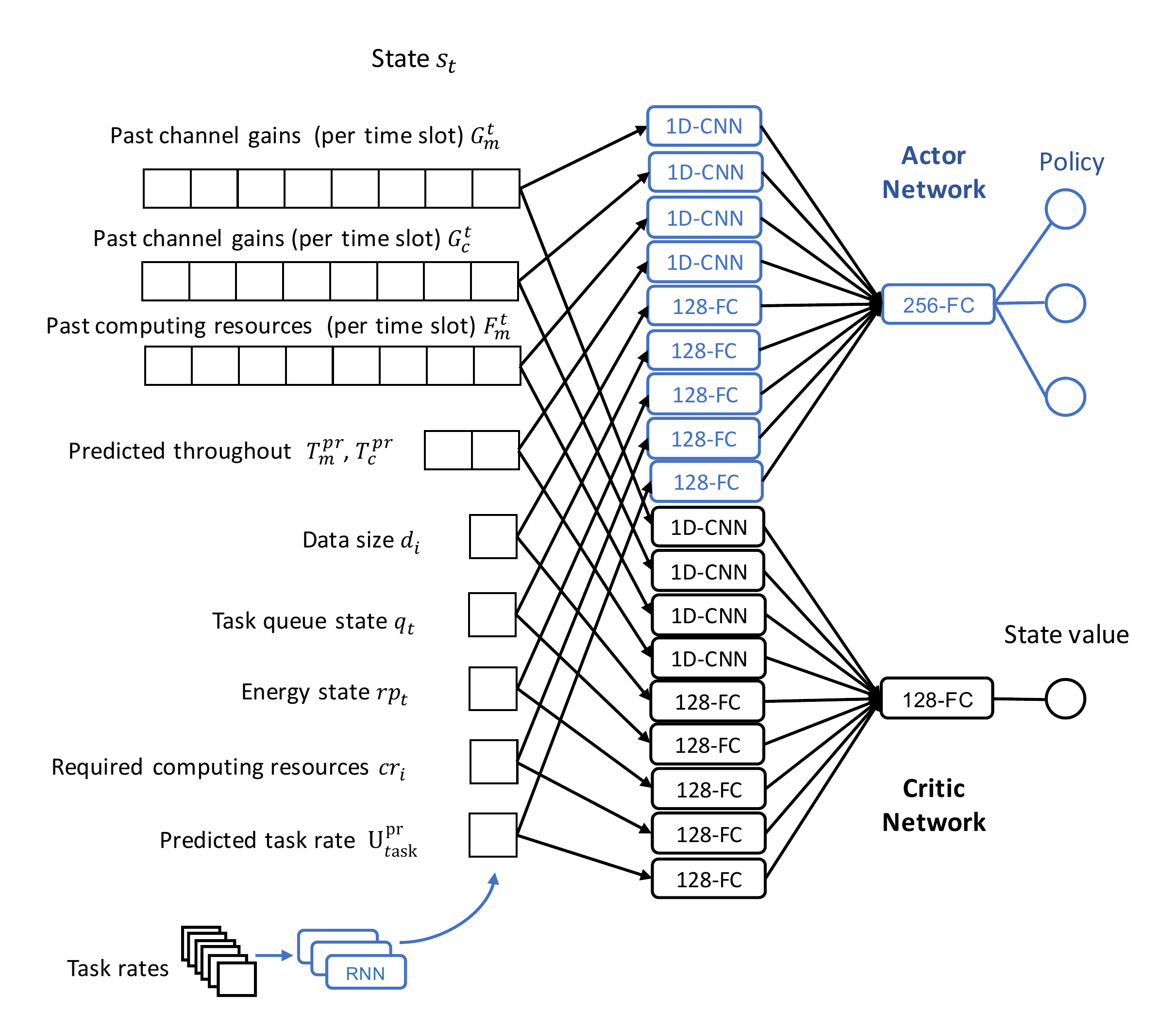}}
	\caption{Neural network architecture of task offloading for DT empowered IoV.}
	\label{fig:NN}
\end{figure}

In terms of network architecture, as shown in Fig.~\ref{fig:NN}. It consists of two parts: actor network and critic network, and they all consist of convolutional neural network (CNN) and fully connected (FC) networks. The input of the neural network is the state of the environment observed by the agent. The output of actor network and critic network are offloading decision and value state, respectively. In order to make better use of state information, we connect the vectors in the state to the convolutional layer, and connect the scalar to the fully connected layer. Note that the wireless channel is dynamically changing, and there is a potential correlation between adjacent channel gains. In addition, digital twins can obtain past channel gains. Therefore, we use RNN to learn the characteristics of channel gains and predict the average throughput for a period of time in the future. Particularly, the arrival of tasks is time-dependent, so we use the past task rates as the input of RNN to predict the arrival of future tasks. 

\subsection{Training Algorithm}
Asynchronous advantage actor-critic is used to train our model. It is based on actor-critic framework and updates parameters asynchronously. The framework consists of critic network and actor network.

\textbf{Actor network.} The primary objective of the actor network is to find policy  $\pi_{\theta }$ that maximize the expected accumulated reward. Noted that the output of the actor network is the distribution of actions (i.e., $\pi _{\theta }\left( s_{i}\right) $) corresponding to the current state $i$ under policy $\pi_{\theta }$. Therefore, we use a policy gradient (PG) method to train the policy. The gradient of the objective with parameter $\theta$ is given by
\begin{equation}
	\nabla _{\theta }E_{\pi _{\theta }}\left[ \sum\limits_{l=0}^{\infty }\gamma
	^{l}r_{l}\right] =E_{\pi _{\theta }}\left[ \nabla _{\theta }\log \pi
	_{\theta }\left( s,a\right) A^{\pi _{\theta }}\left( s,a\right) \right],
\end{equation}
where $A^{\pi _{\theta }}\left( s,a\right) $ is the advantage function, which is computed by the critic network, and it indicates how much a specific action is better than the average action taken under the policy~\cite{mao2017neural}. The updating process of the actor network with parameter $\theta$ follows 
\begin{equation}
	\begin{array}{lll}
		\theta =\theta  & + & \alpha ^{actor}\sum\limits_{i}\nabla _{\theta }\log
		\pi _{\theta }\left( s_{i},a_{i}\right) A\left( s_{i},a_{i}\right)  \\ 
		& + & \phi \sum\limits_{i}\nabla _{\theta }H\left( s_{i},a_{i}\right) 
	\end{array}\label{equ:actor_update} 
\end{equation} 
where $\alpha^{\textrm{actor}}$ is the learning rate of the actor network, and $H\left( \cdot \right) $ is the entropy of the policy at each state. $\phi $ is regularization parameters which controls the strength of the entropy regularization term.

\begin{figure}[t]
	\centerline{\includegraphics[scale=0.2]{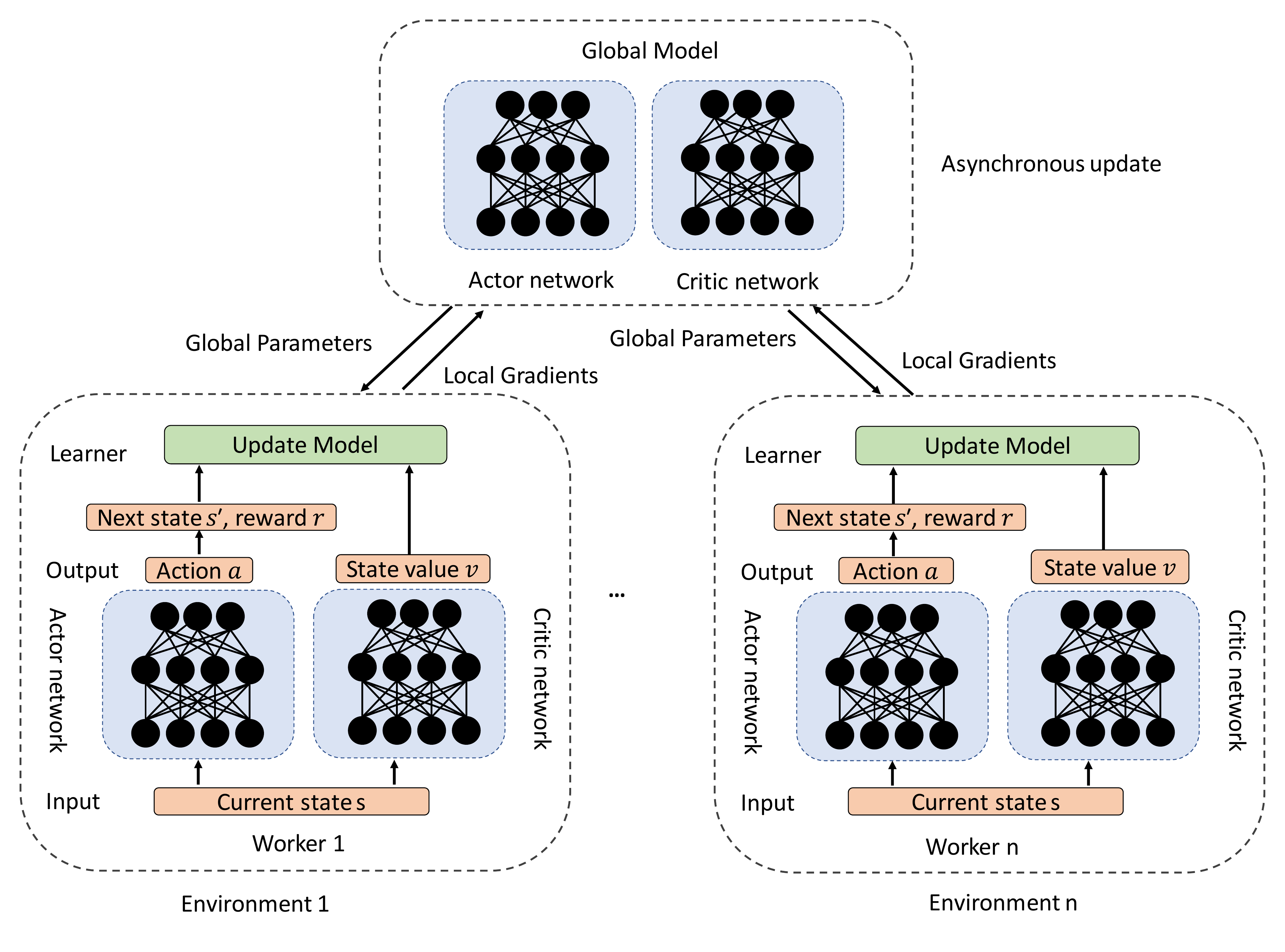}}
	\caption{Training offloading model using A3C.}
	\label{fig:train}
\end{figure}
\textbf{Critic network.}
The critic network uses value-based method to evaluate the current state, denoted as  the value function $V^{\pi _{\theta }}\left( s\right) $. The value function is the expected total reward starting from state $s$ and under the policy $\pi_{\theta }$. Let $\theta_{v}$ denote the parameters of the critic network, the updating of the parameters $\theta_{v}$ can be defined as 
\begin{equation}
	\theta _{v}=\theta _{v}+\alpha ^{\textrm{critic}}\sum\limits_{i}\left(
	R_{i}^{m}-V^{\pi _{\theta }}\left( s_{i};\theta _{v}\right) \right) ^{2},\label{equ:update_critic}
\end{equation}
where $\alpha ^{\textrm{critic}}$ is the learning rate for the critic network, and $V^{\pi _{\theta }}\left( \cdot ;\theta _{v}\right) $ is the estimate of $V^{\pi _{\theta }}\left( \cdot \right) $. $R_{i}^{m}$ denotes $m$-step accumulated discounted reward of the agent from state $i$ following the standard Temporal Difference (TD) method, which is given by
\begin{equation}
	R_{i}^{m}=\sum\limits_{l=0}^{m-1}\gamma _{i+l}^{l}+\gamma ^{m}V^{\pi
		_{\theta }}\left( s_{i+m};\theta _{v}\right) \label{equ:accumulated_reward},
\end{equation}
where $\gamma \in \left[ 0,1\right] $ is the discount factor. Accordingly, the advantage function in (\ref{equ:actor_update}) is defined as
\begin{equation}
	\begin{array}{lll}
		A\left( s_{i},a_{i}\right)  & = & \sum\limits_{l=0}^{m-1}\gamma
		_{i+l}^{l}+\gamma ^{m}V^{\pi _{\theta }}\left( s_{i+m};\theta _{v}\right)
		-V^{\pi _{\theta }}\left( s_{i};\theta _{v}\right)  \\ 
		&  &  \\ 
		& = & R_{i}^{m}-V^{\pi _{\theta }}\left( s_{i};\theta _{v}\right) 
	\end{array}%
\end{equation}

In our scenario, the model is trained in a dynamic environment, the correlation between adjacent states makes it difficult or even impossible for the agent to converge. We use parallel training method based on A3C~\cite{A3C,Mu2105:AMIS}. It uses an asynchronous update mechanism to solve the correlation between adjacent states, making training more efficient. Fig.~\ref{fig:train} shows the A3C training framework with multiple parallel workers. The framework consists of a global model and $n$ local workers working asynchronously, and each worker runs a copy of the model independently. Workers regularly push the gradient of local parameters into the global model, and extract the latest parameters from the global model to update the local model. For the global model, it will integrate the gradient received from each worker and update the global parameters. The  training algorithm of task offloading using A3C for each worker is shown in Algorithm~\ref{algorithm}.
\begin{algorithm}
	\caption{Training Algorithm of Task offloading Based on A3C for each worker}
	\label{alg:DPA}
	\begin{algorithmic}
		\STATE{// network parameters for global network: $\theta ^{\textrm{g}}$, $\theta _{\textrm{v}}^{\textrm{g}}$}
		\STATE{// network parameters for each worker: $\theta ^{\textrm{w}}$, $\theta_{\textrm{v}}^{\textrm{w}}$}
		\STATE{Initialize learning rate $\alpha ^{\textrm{actor}}$ and $\alpha ^{\textrm{critic}}$, worker step counter $i=1$, global counter $I=0$, maximum number of steps in each episode $m$}

		\REPEAT 
		\STATE{Gradient initialization: $d\theta ^{\textrm{g}}=0$, $d\theta _{\textrm{v}}^{\textrm{w}}=0$}
		\STATE{Synchronize parameters: $\theta ^{\textrm{w}}=\theta ^{\textrm{g}}$, $\theta_{\textrm{v}}^{\textrm{\textrm{w}}}=\theta _{\textrm{v}}^{\textrm{g}}$}
		\STATE{Observe the state  $s_{i}$ of the vehicle}
		\STATE{$i^{^{\prime }}=i$ }
		\REPEAT 
		\STATE{Take action $a_{i}$ according to policy $\pi _{\theta }$, compute the reward $r_{i}$ according to (\ref{equ:reward}), and get the next state $s_{i+1}$}
		\STATE{$i=i+1$}
		\STATE{$I=I+1$}
		\UNTIL{$i-i^{^{\prime }}==m$ or terminal $s_{i}$} 
		
		\STATE{$R=\left\{ 
			\begin{array}{c}
				0, \\ 
				v\left( s_{i};\theta _{\textrm{v}}^{\textrm{w}}\right) ,
			\end{array}%
			\right. 
			\begin{array}{c}
				\textrm{if} \ \textrm{terminal} \ s_{i}\\ 
				\textrm{otherwise}
			\end{array}
			$}
		\FOR{$j\in \left\{ i-1,...,i^{^{\prime }}\right\} $}
		\STATE{$R=r_{j}+\gamma R$}
		\STATE{$d\theta ^{\textrm{g}}=d\theta ^{\textrm{g}}+\nabla _{^{\theta ^{\textrm{w}}}}\log \pi _{\theta
				^{\textrm{w}}}\left( a_{j},s_{j}\right) \left( R-V^{\pi _{\theta }}\left(
			s_{j;}\theta _{\textrm{v}}^{\textrm{w}}\right) \right) $}
		
		\STATE{$d\theta _{v}^{\textrm{g}}=d\theta _{\textrm{v}}^{\textrm{g}}+\frac{\partial \left( R-V^{\pi _{\theta}}\left( s_{j};\theta _{\textrm{v}}^{\textrm{w}}\right) \right) ^{2}}{\partial \theta _{\textrm{v}}^{\textrm{w}}}$}
		\ENDFOR
		
		\STATE{Update $\theta ^{\textrm{g}}$ and $\theta^{\textrm{g}}_{\textrm{\textrm{v}}}$ using $d\theta^{\textrm{\textrm{g}}}$ and $d\theta^{\textrm{g}}_{\textrm{v}}$ respectively } 
		
		\UNTIL{$I>I_{max}$} 
		
	\end{algorithmic}\label{algorithm}	
\end{algorithm}

\section{Simulations} 
In this section, we discuss the experimental setting, baselines, and the performance of the proposed algorithm.
\subsection{Experiment Setup}
Our algorithm and network architecture are implemented using TensorFlow 2.0 and Keras. In our experiment, the available devices are 1 Intel Core i9 10980xe CPU, which has 16 cores, and we allow 64GB of RAM for model training. In addition, the neural network is trained on Nvidia GeForce RTX 2080ti GPU with 11GB of memory.
\subsubsection{Environment settings}
We simulate a scenario depicted in Fig.~\ref{fig:DT_offloading} where vehicles are driving along a road covered by BSs and RSUs. Specifically, We simulated the task offloading model of a single vehicle. As mentioned earlier, the interaction between vehicles is only related to their shared use of MEC computing resources. Therefore, we express the dynamic changes of MEC computing resources as the mutual influence between multiple vehicles. In our experiment, the length of each time slot is set as 0.2s and the distance between RSUs is set to 50 meters (i.e., $L=50$). We set $v=10 m/s$ as default, and the length of task queue is set to 1000 (i.e., $\left\Vert Q\right\Vert =1000$). We set $T_{i}^{\max }=20$ for all tasks. The data size $d_{i}$ and required computing resources $cr_{i}$ for task $i$ are both sampled from uniform distributions within $\left[ 0.1,2.5\right]$ MB and $\left[ 1,10\right] $ Gigacycles, respectively. The task arrival rate is selected from $\left[ 0.1,0.9\right]$ per time slot based on Markov Chain, which detailed in Appendix~\ref{append:task_pred}. The transmission power of each vehicle is set to $1.25W$, and the carrier frequency is set to $2$ GHz. Besides, the wireless bandwidth of MEC server and cloud are set to 10 MHz and 5 MHz, and the noise power is set to -114 dBm. The computation capacity of vehicle and cloud server are set to $2$ and $20$ Gigacycles per time slot. The cost of renting one unit computation resource of cloud server $\varpi _{com}$ and the price factor $\mu $ are set to 3 and 1 respectively, and the communication cost $\varpi _{tr}$ is set to 1. The power for local vehicle $p_{n}$ can be obtained by $p_{n}=\zeta \left( f_{n}^{\textrm{l}}\right) ^{\tau }$. Here, we set $\zeta=1$ and $\tau=3$ according to~\cite{dinh2017offloading}. In addition, the computation capacity of MEC server is sampled from uniform distributions within $\left[ 2,10\right]$ Gigacycles per time slot. In particular, we ignore the communication delay between the digital twin and the vehicle. 
\subsubsection{Training settings}
The structure of neural network for computing task offloading in our work is shown in Fig~\ref{fig:NN}. For both the actor network and critic network, the first hidden layer consists of five FC layers with 128 neurons, and four CNN layers with 64 filters of size 4x4, stride 1. The size of second hidden layer for actor network and critic network are set to 256 and 128, respectively. Finally, the output layer of actor network contains 3 units corresponding to actions, and critic network contains 1 units representing the value of current state. For both the actor network and critic network, relu6 is used as activation function, and RMSProp is selected as the optimizer. In addition, we set the number of past time slots $u$ (i.e., the length of $F_{m}$) for MEC's computation capacity to 5, and set the number of past time slots $j$ (i.e., the length of $G_{m}$ and $G_{c}$) of gains for MEC server and cloud to 50. Other parameters used for neural network training are shown in Table~\ref{table:parameters}. The throughput from vehicle to MEC server and cloud is given by two prediction modules respectively and the task arrival rate is given by task prediction module. Further details about the prediction module settings for Digital twin agent can be found in Appendix \ref{append:throu_pred}.
\begin{figure*}
	\centering
	\begin{subfigure}[t]{0.28\textwidth}
		\centering
		\includegraphics[width=\textwidth]{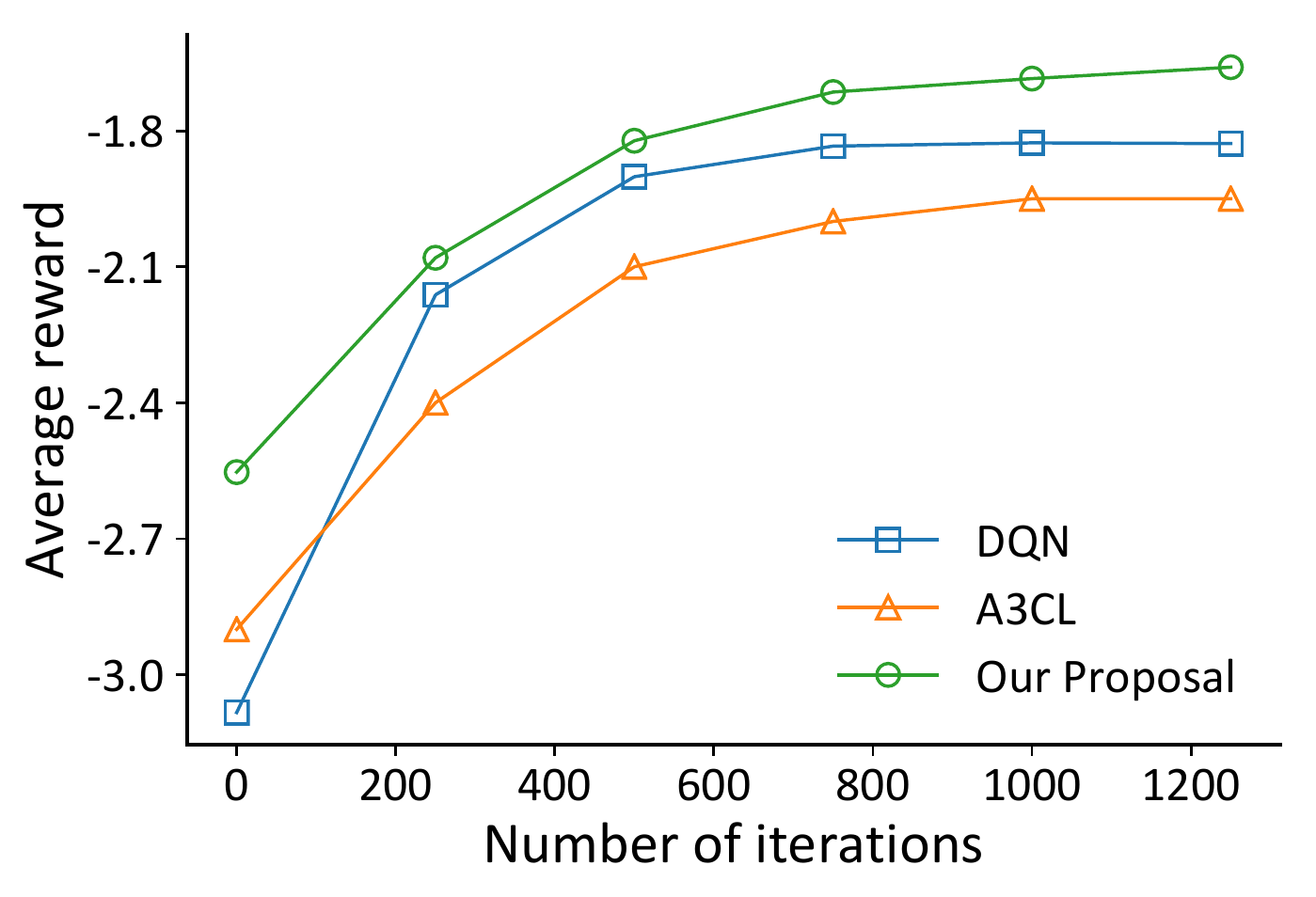}
		\caption{Training curve of algorithm}
		\label{fig:reward}
	\end{subfigure}
	\hfil
	\begin{subfigure}[t]{0.28\textwidth}
		\centering
		\includegraphics[width=\textwidth]{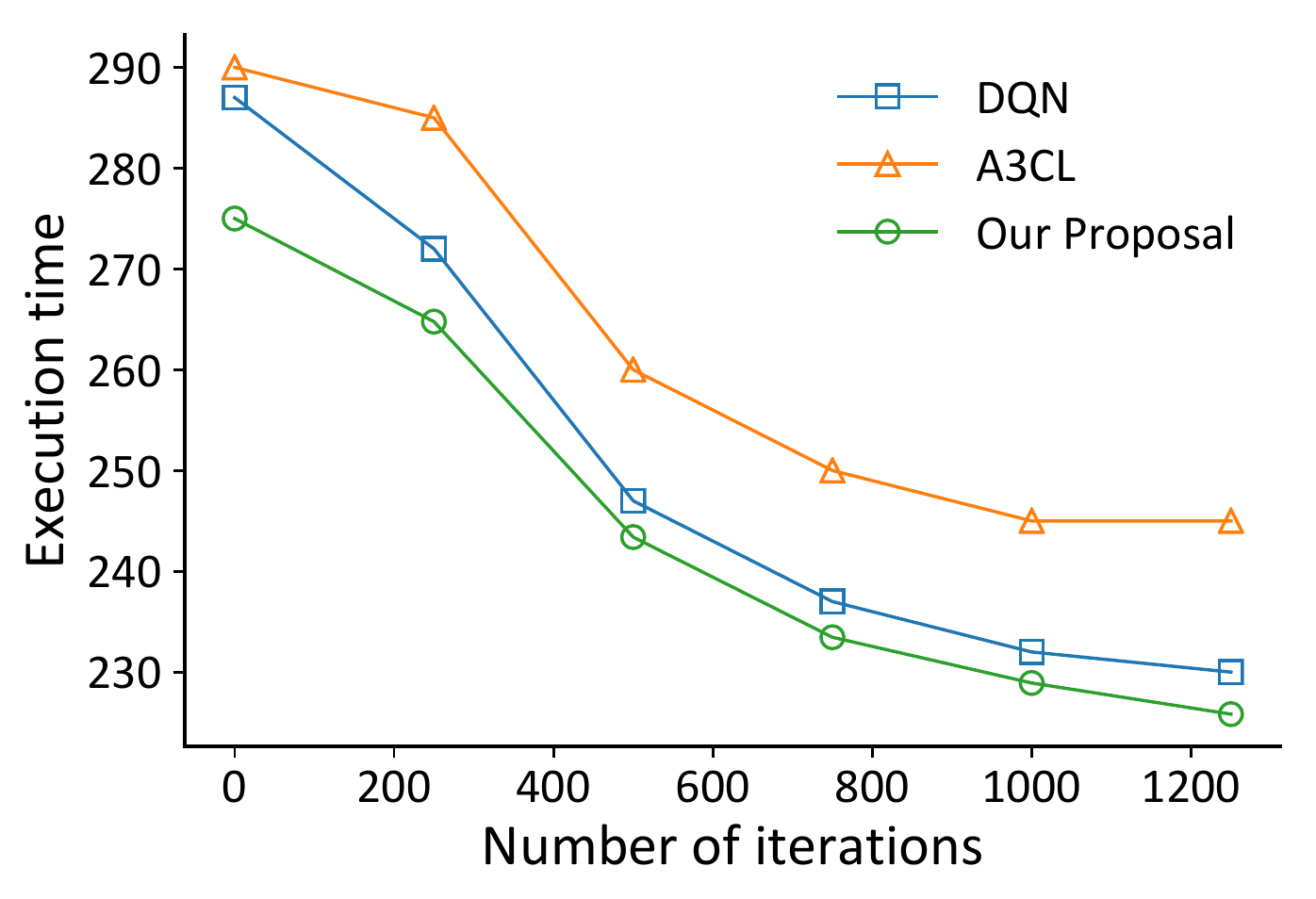}
		\caption{Execution time curve}
		\label{fig:time}
	\end{subfigure}
	\hfil
	\begin{subfigure}[t]{0.28\textwidth}
		\centering
		\includegraphics[width=\textwidth]{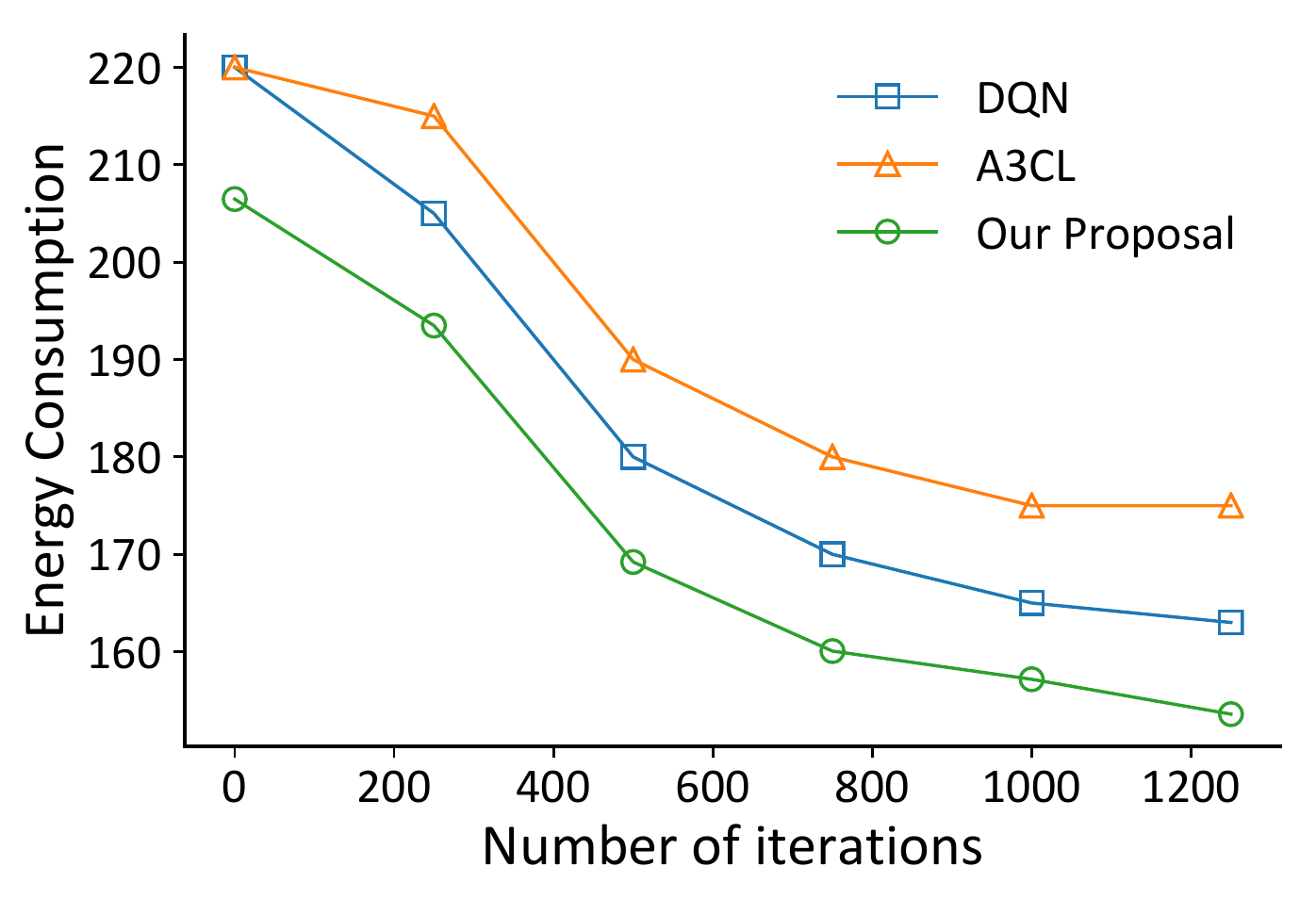}
		\caption{Energy consumption curve}
		\label{fig:energy}
	\end{subfigure}
	\caption{Convergence performance of algorithms in training.}
	\label{fig:training}
\end{figure*}

\begin{table}[]\caption{Training Hyperparameters}
	\centering
	\begin{tabular}{|c|c|c|ll}
		\cline{1-3}
		Parameter & Description                     & Value &  &  \\ \cline{1-3}
		$\alpha^{\textrm{actor}}$          & Learning rate of actor network  & $10^{-5}$     &  &  \\ \cline{1-3}
		$\alpha^{\textrm{critic}}$          & Learning rate of critic network & $10^{-5}$     &  &  \\ \cline{1-3}
		$\phi $         & Regularization parameters       & $0.01$     &  &  \\ \cline{1-3}
		$\gamma$          & Discount factor                 & $0.9$     &  &  \\ \cline{1-3}
		$n$         & Number of works in training     & $32$     &  &  \\ \cline{1-3}
		$\xi _{1}$         & Weight factor for execution time                & $0.4$     &  &  \\ \cline{1-3}
		$\xi _{2}$         & Weight factor for energy consumption       & $0.4$     &  &  \\ \cline{1-3}
		$\xi _{3}$         & Weight factor for renting       & $0.2$     &  &  \\ \cline{1-3}
		$\eta $         & Penalty coefficient for time                & $5$     &  &  \\ \cline{1-3}
		$\psi $     & Penalty coefficient for overflow            & $5$    &  &  \\ \cline{1-3}
		$v$         & Scaling parameter for reward       & $0.005$    &  &  \\ \cline{1-3}
	\end{tabular}\label{table:parameters}
\end{table} 
\begin{figure*}
	\centering
	\begin{subfigure}[t]{0.23\textwidth}
		\centering
		\includegraphics[width=\textwidth]{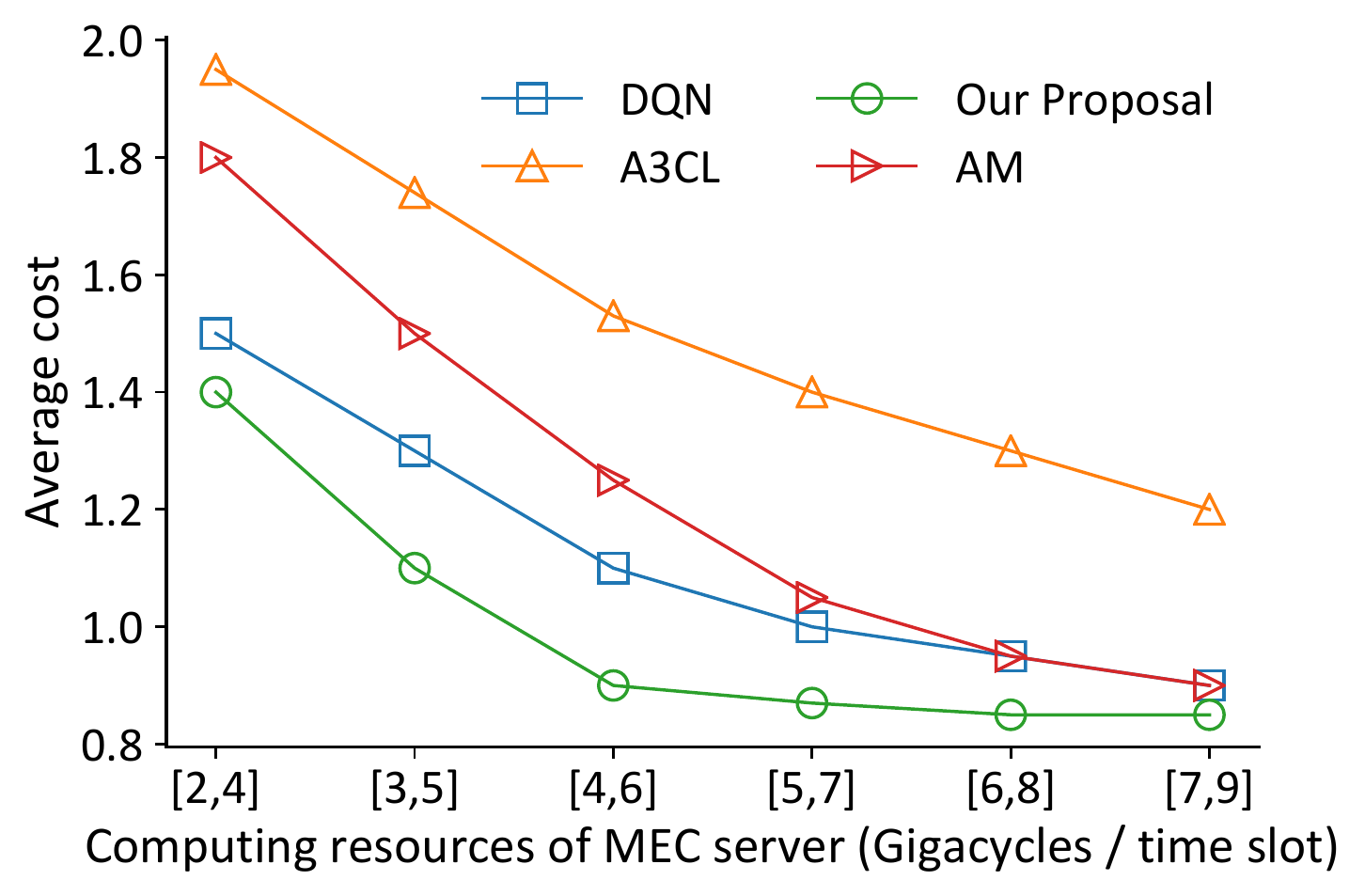}
		\caption{Average cost of different algorithm}
		\label{fig:static_average_cost}
	\end{subfigure}
	\hfil
	\begin{subfigure}[t]{0.23\textwidth}
		\centering
		\includegraphics[width=\textwidth]{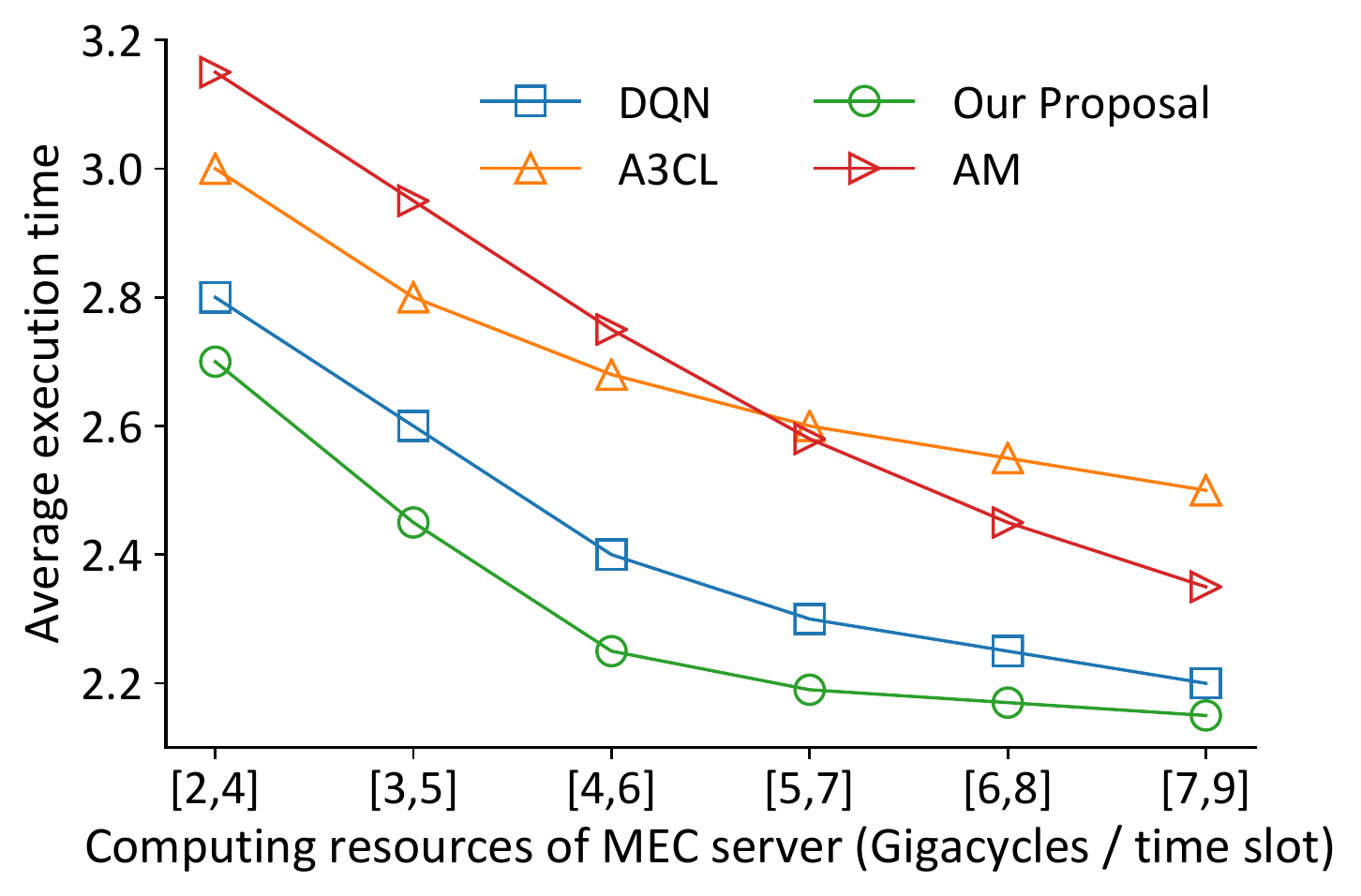}
		\caption{Execution time}
		\label{fig:static_time}
	\end{subfigure}
	\hfil
	\begin{subfigure}[t]{0.23\textwidth}
		\centering
		\includegraphics[width=\textwidth]{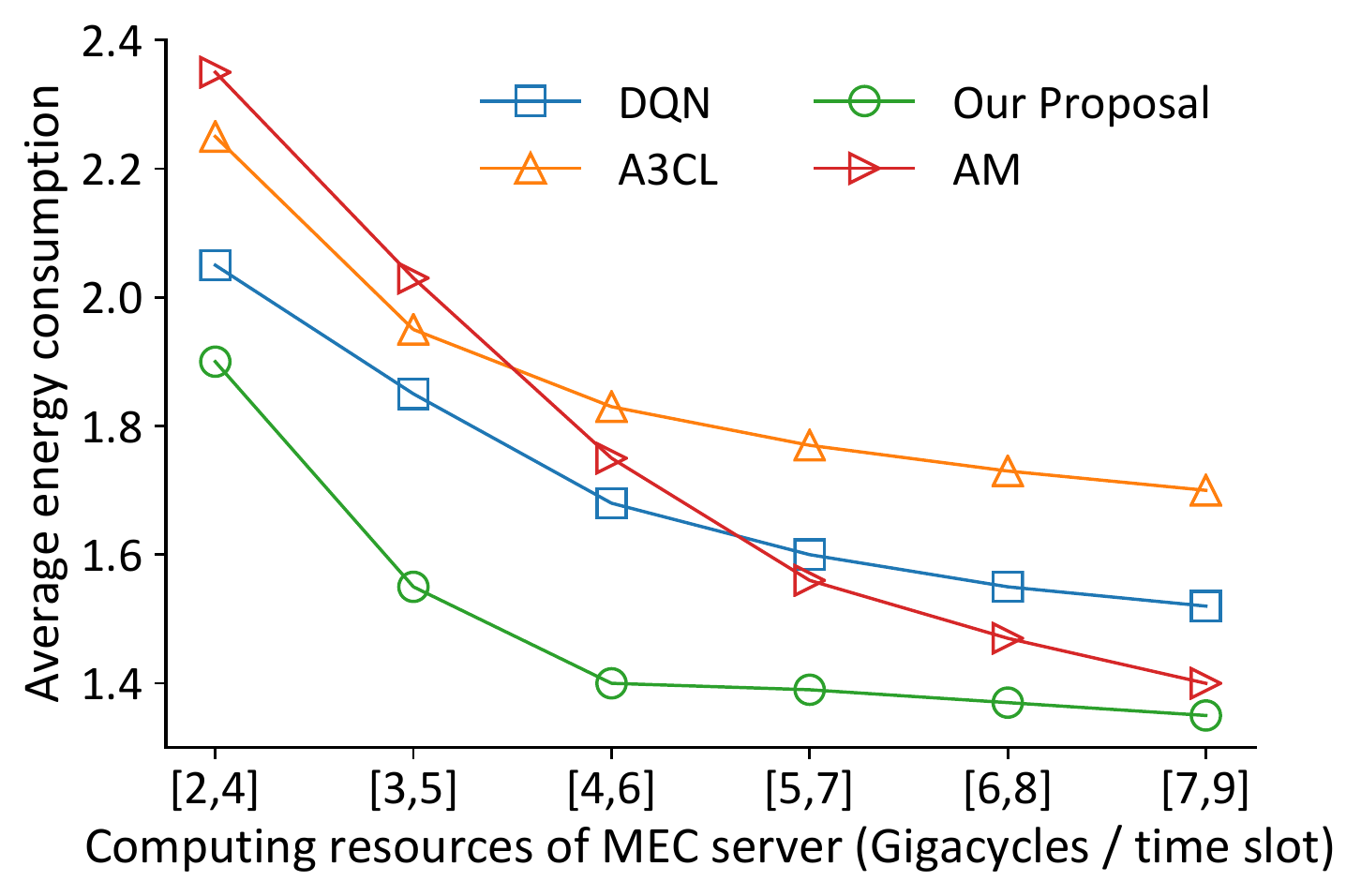}
		\caption{Energy consumption}
		\label{fig:static_energy}
	\end{subfigure}
	\hfil
	\begin{subfigure}[t]{0.23\textwidth}
		\centering
		\includegraphics[width=\textwidth]{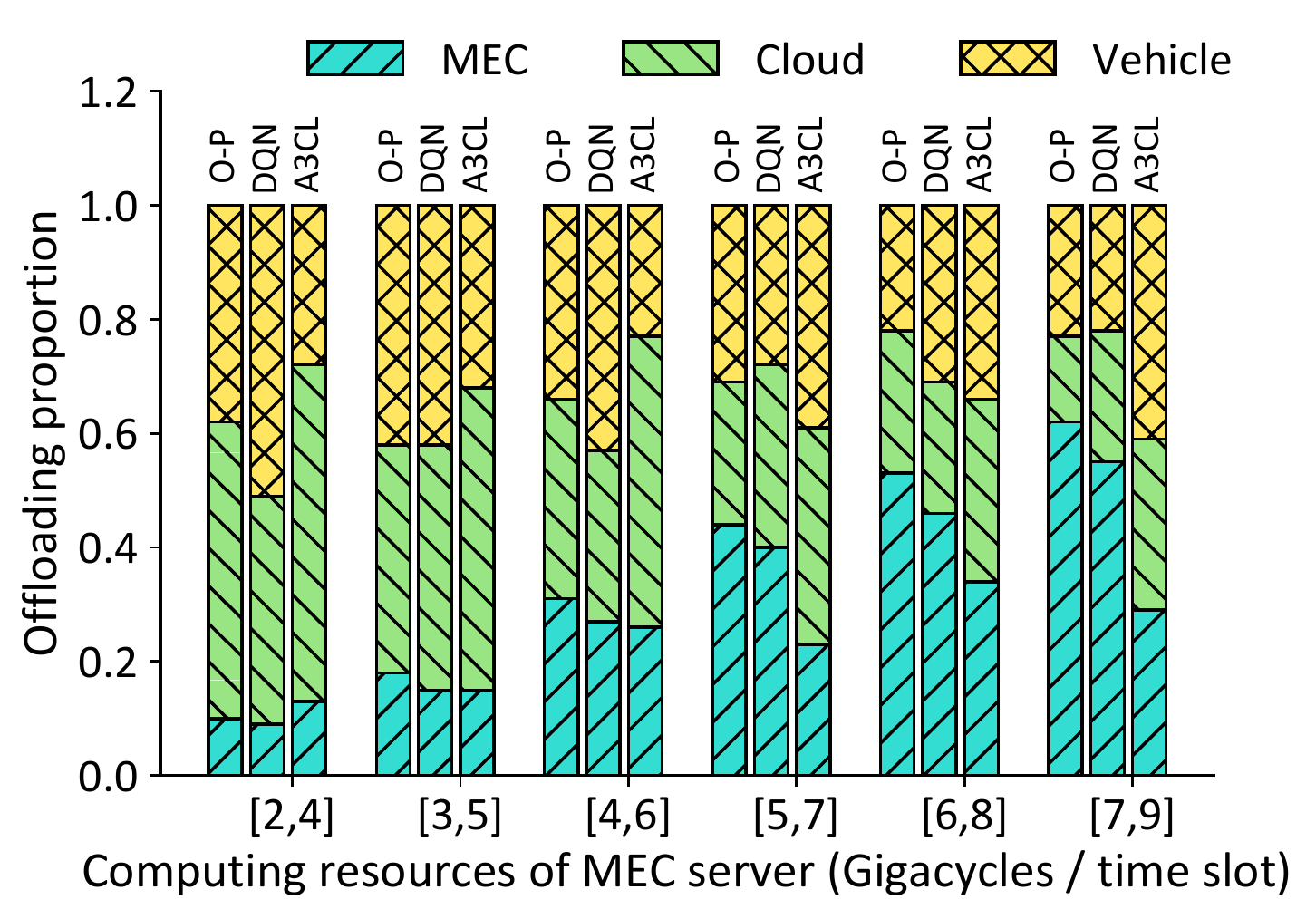}
		\caption{Offloading proportion}
		\label{fig:static_ratio}
	\end{subfigure}
	\caption{Performance comparison under different computing resources of MEC server.}
	\label{fig:static_MEC}
\end{figure*}

\subsection{Baselines}\label{baselines}
\subsubsection{All local execution (AL)}
In this scenario, all tasks are executed on local vehicles.
\subsubsection{All offloaded to MEC (AM)}
At any state $s_{t}$, regardless of the wireless conditions and the computing resources of the MEC server, all tasks are offloaded to the MEC server for execution.
\subsubsection{All offloaded to cloud (AC)}
Similar to AM, all tasks are offloaded to cloud server for execution, regardless of the wireless conditions and the renting for using cloud.
\subsubsection{Rondom choice (RC)}
We randomly choose one of three computing units (i.e., local vehicle, MEC server, and cloud) with the same probability to execute the task. 

\subsubsection{Deep Q-Learning based algorithm (DQN)}
Most of the existing work is to use Deep Q-Learning algorithm to solve computation offloading problems. In order to show the advantages of our proposal, we design a task offloading model based on DQN for comparison. In this scenario, DQN based algorithm including Q-Network and Target Q-Network is used to train task offloading model. We set the same training parameters in DQN as our algorithm. 

\subsubsection{A3C in local vehicle (A3CL)}
Different from our proposal in digital twins, the scheduling module (i.e., DRL agent) will be deployed in local vehicle. In this scenario, due to the lack of global information (i.e., the past channel gains and MEC information), the vehicle can only make offloading decisions based on the current state information (e.g., current channel information and MEC states). This method is used to show the performance of our algorithm without the support of digital twins.

\begin{figure}[t]
	\centerline{\includegraphics[scale=0.5]{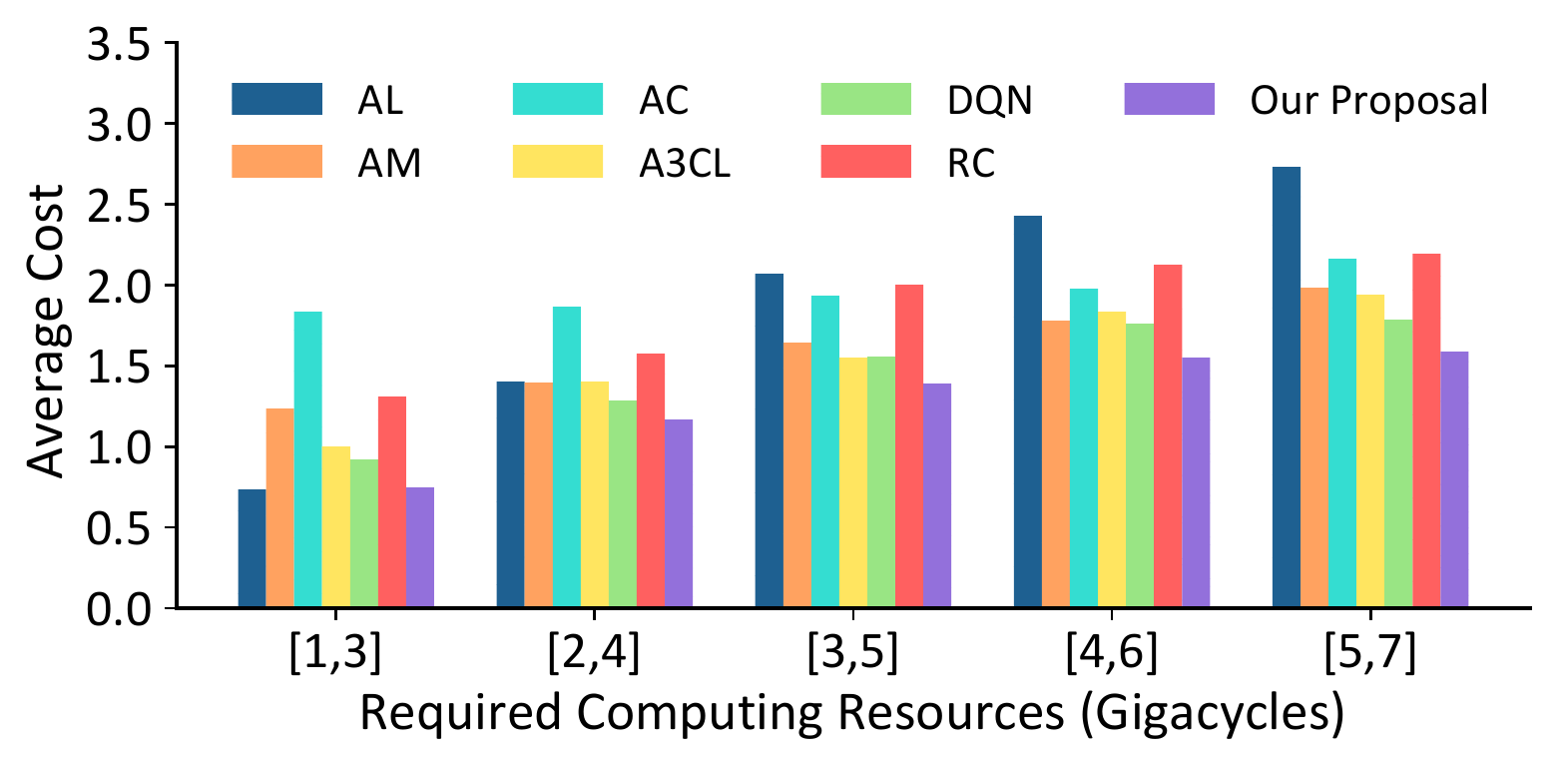}}
	\caption{Average cost under different computing resources required by tasks.}
	\label{fig:static_required_r}
\end{figure}

\begin{figure}[tbp]
	\centerline{\includegraphics[scale=0.5]{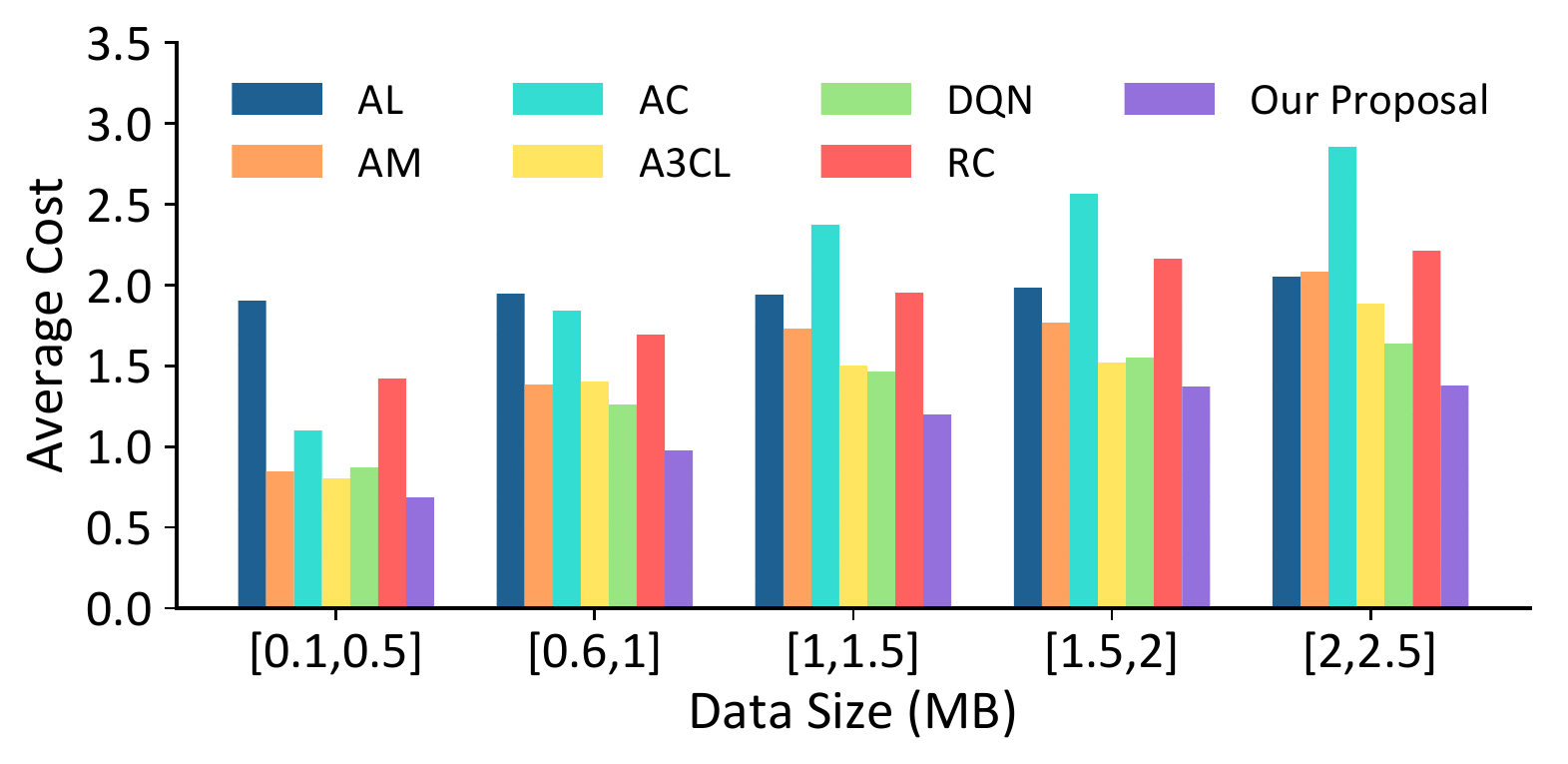}}
	\caption{Average cost under different data size of tasks.}
	\label{fig:static_data_size_r}
\end{figure}

\begin{figure*}
	\centering
	\begin{subfigure}[b]{0.32\textwidth}
		\centering
		\includegraphics[width=\textwidth]{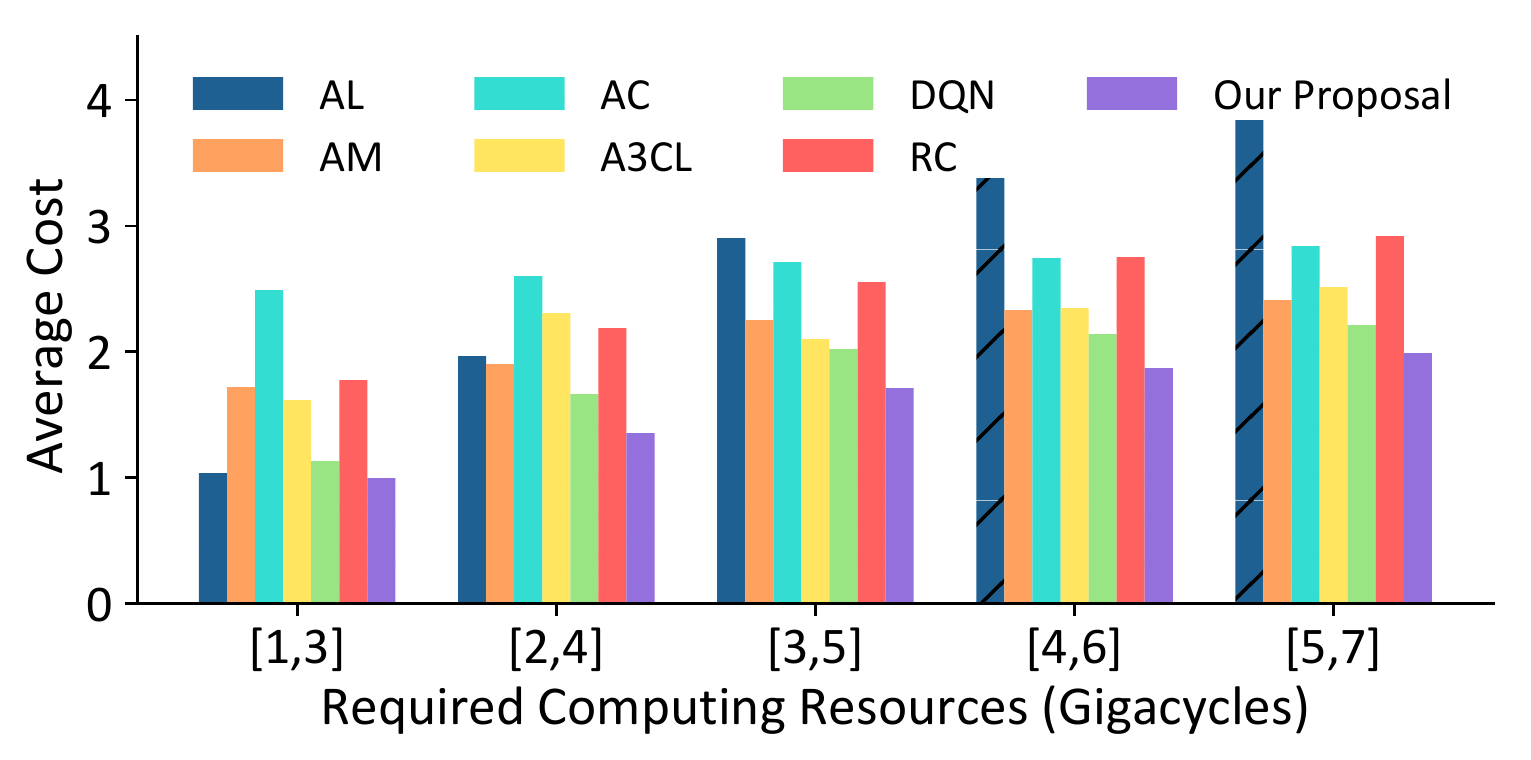}
		\caption{Required computing resources by tasks}
		\label{fig:dynamic_required_r}
	\end{subfigure}
	\hfil
	\begin{subfigure}[b]{0.32\textwidth}
		\centering
		\includegraphics[width=\textwidth]{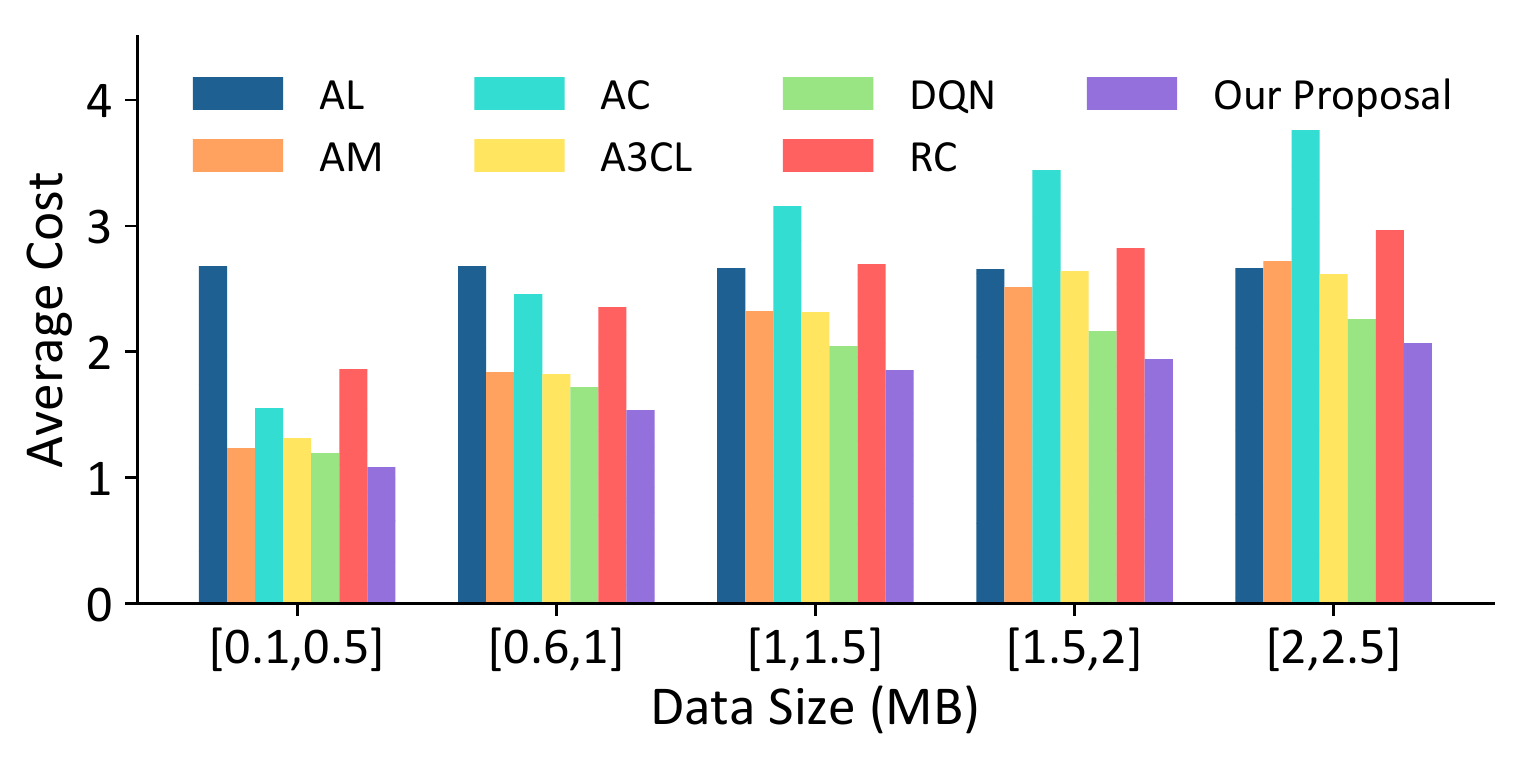}
		\caption{Data size of tasks}
		\label{fig:dynamic_data_size_r}
	\end{subfigure}
	\hfil
	\begin{subfigure}[b]{0.32\textwidth}
		\centering
		\includegraphics[width=\textwidth]{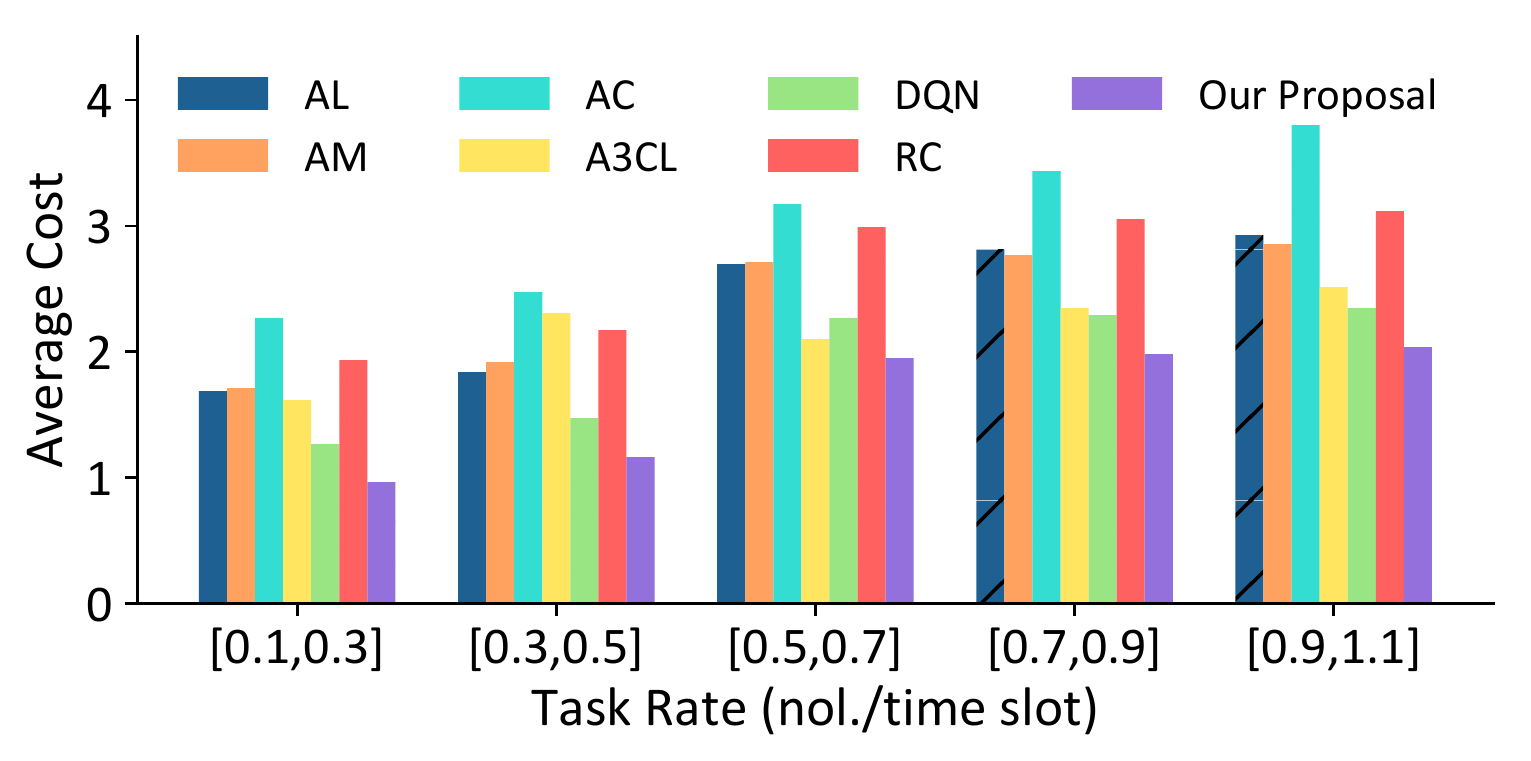}
		\caption{Task rate}
		\label{fig:dynamic_task_r}
	\end{subfigure}
	\caption{Comparison of average cost in dynamic queue scenario.}
	\label{fig:dynamic_scenario}
\end{figure*}

\subsection{Training Performance}
We train each algorithm 50 times and take the average of the reward. The average reward  comparison after being scaled is shown in Fig.~\ref{fig:reward}. It is clear that the our algorithm can always converge to a higher level than the other algorithms after a small number of iterations because our algorithm can quickly and effectively find the optimal offloading policy. The convergence rate and reward of DQN proposed in the previous work are inferior to our proposal. Note that the convergence speed and the reward obtained of A3CL are the worst. The reason is that without the assistance of the digital twin, the vehicle lacks global and historical information about the environment when making offloading decisions. This will make the vehicle can only make offloading decisions based on one-sided information, leading to a sub-optimal solution.

In order to verify the effectiveness of our proposed algorithm, we consider individually the task execution time and energy consumption. The cost of execution time and energy consumption are shown in Fig.~\ref{fig:time} and Fig.~\ref{fig:energy} respectively. It can be seen from the figure the three algorithms optimize task execution delay and energy consumption at the same time. In particular, compared with DQN and A3CL, our algorithm has the best convergence performance in both execution time and energy consumption. 
\subsection{Performance in Static Queue Scenario}
In this scenario, we randomly initialize the task queue and assume that no new tasks arrive after initialization (i.e., $\lambda=0$). Based on this, a complete training episode starts from the initialization of the task queue until there are no tasks in the queue. We evaluate the performance of our algorithm through comparisons with the baselines mentioned in Section~\ref{baselines}.

In our scenario, the available resources of the MEC server are affected by other vehicles (i.e., $f_{k}^{t}$ is a function of time). In order to investigate the impact of changes in the available computing resources of the MEC server on the offloading policy, we conduct simulations under different MEC computing resources.  Fig.~\ref{fig:static_MEC} shows the performance comparison between three baselines and our proposal under different computing resources of MEC server. In Fig.~\ref{fig:static_average_cost} , the average cost of tasks is illustrated. We can see that the performance of all algorithms improves as the computing resources increase. Note that the cost of all algorithms decrease faster when computing resources increases from [2,4]\footnote{[2,4] means that the available computing resources of MEC server is sampled from uniform distributions within [2,4] Gigacycles per time slot.} GHz to [4,6] GHz than other cases. This is because when the computing resources of MEC server is too low, tasks can only be executed on local vehicle or the cloud server, resulting in higher cost of tasks. As the computing resources increases, more tasks are offloaded to MEC server for execution, resulting in a rapid decrease in cost. However, when the computing resources continues to increase, the cost of tasks will not be affected by the computing resources (i.e., time and energy consumption of transmission are the main factors that affect the cost of tasks). In particular, A3CL lacks the assistance of digital twins, resulting in higher cost of tasks. 

The comparison of average execution time is shown in Fig.~\ref{fig:static_time}. From the figure, the execution time of all algorithm decrease with the increasing of computing resources of MEC server. Note that our algorithm can reach the lower execution time than other baselines. Fig.~\ref{fig:static_energy} compares the energy consumption of the four algorithms. It is observed that the change in energy consumption is similar to the change in execution time, because energy consumption is related to time (i.e., energy consumption is the product of power and time). Note that when the computing resources of MEC server increase to [7,9] GHz, the energy consumption of AM reduce to close to that of our proposal. This is because when the computing resources increasing, more tasks will be offloaded to the MEC server, and the energy consumption is mainly generated by task transmission (usually the energy consumption of task transmission is less than the that of task execution). Fig.~\ref{fig:static_ratio} shows the offloading proportion for tasks under different computing resources of MEC server. For the convenience of drawing, we replace "Our Proposal" with "O-P". As shown, the proportion of MEC server increases with the increase of computing resources. However, the proportion of the cloud decreases because the cost of tasks (i.e., execution time, energy consumption, and rent) executed on the cloud at this time exceeds the cost of tasks executed on MEC server. In this case, the digital twin will reduce the number of tasks offloaded to the cloud to reduce cost. In addition, for AM, all tasks are offloaded to the MEC server (i.e., the offloading proportion of AM is $1.0$ while the offloading proportion of other algorithms are all equal to $0$), which we do not show in the figure.

To show the performance of our algorithm, we implement all the baselines under different computing resources required by tasks and data size of tasks, respectively. The average cost comparison versus required computing resources for tasks is shown in Fig.~\ref{fig:static_required_r}. We can see that AL, AM, AC, and RC always have a high cost because they cannot adapt to complex environmental conditions. Especially when the task requires a lot of computing resources, AL performs very badly due to the limited computing power of the vehicle. Fortunately, our proposal and DQN can find the optimal offloading policy, and our algorithm has a lower cost than DQN. Fig.~\ref{fig:static_data_size_r} illustrates the average cost comparison of algorithm performance in different data size of tasks. It can be seen that AC gets a high cost when the task size increases. The main reason is that transmitting a large task to the cloud will cause a high latency and more handover occurrence which meaning that offloading tasks to the cloud is not always a good strategy. Note that the our proposal can always achieves the lowest value of cost among other baselines, which shows that our algorithm is more efficient than the other baselines.

We can see from Fig.~\ref{fig:static_average_cost}$\sim$Fig.~\ref{fig:static_energy}, Fig~\ref{fig:static_required_r} and Fig.~\ref{fig:static_data_size_r} that our proposed algorithm can always achieve the lowest average cost, average execution time and average energy consumption, which illustrates the effectiveness of our algorithm in static queue scenario. Next, we investigate the performance of our algorithm in dynamic queue scenario.
\begin{figure}[t]
	\centerline{\includegraphics[scale=0.5]{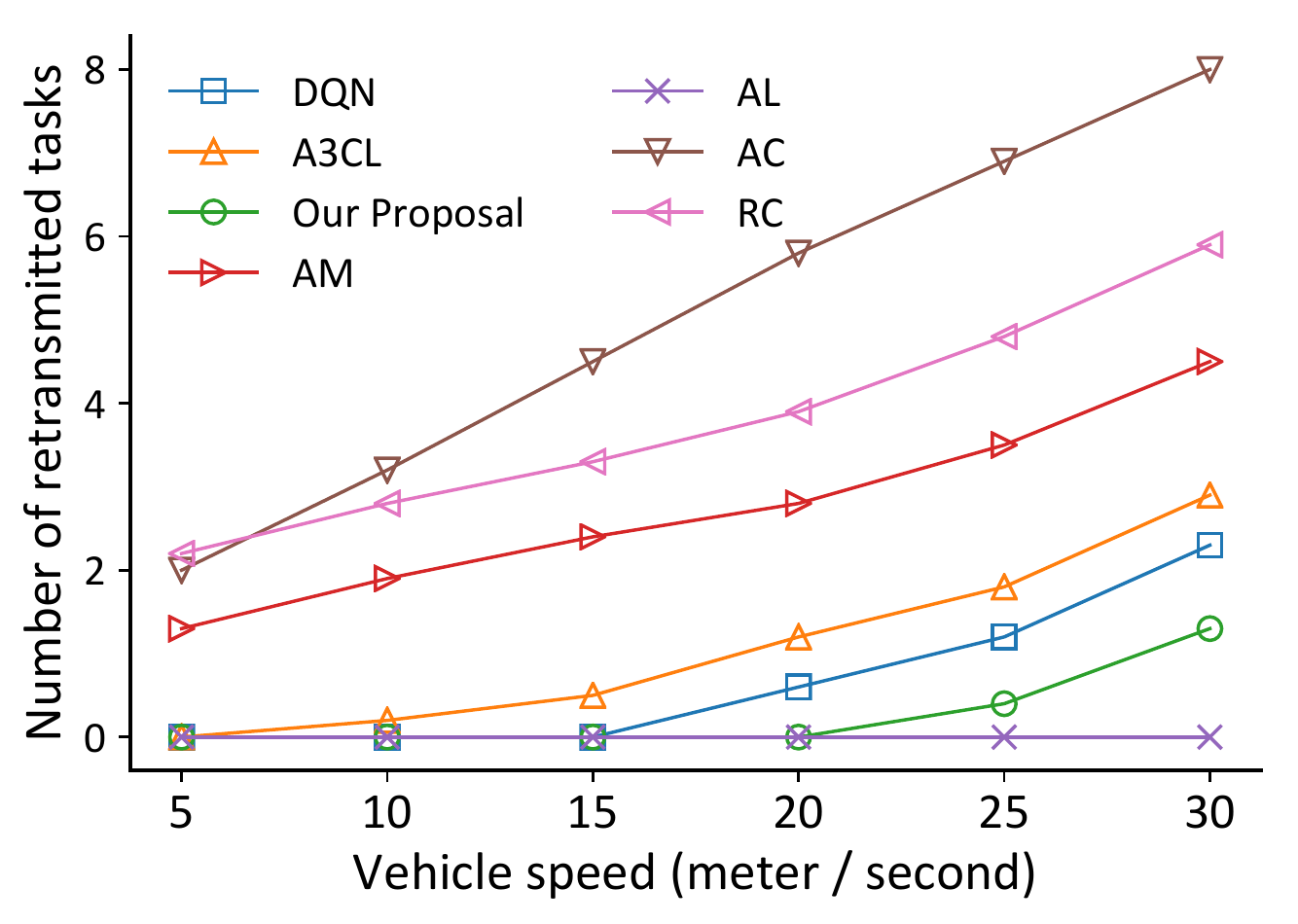}}
	\caption{Average number of re-transmitted tasks in different vehicle speed.}
	\label{fig:diff_v}
\end{figure}
\subsection{Performance in Dynamic Queue Scenario}
Dynamic Queue means that at each time slot, new tasks arrive (i.e., $\lambda >0$), and are appended to the end of the task queue. If no task is executed at this time, the scheduler will take out the task from the head of the task queue for execution according to the offloading actions made by digital twin. We compare
our algorithm with other baselines, and the initial state of all algorithms are set to be identical. In particular, we show the importance of considering task prediction in our algorithm by comparing the performance of our algorithm and A3CL. Note that in a certain time slot $t$, the system has two actions: the vehicle receives the task and puts it in the task queue; the scheduler receives the offloading policy from the digital twin and takes the task out of the queue for execution (if no task is executed at time slot $t$) according to the policy. Therefore, different order of the two actions determines different state (i.e., input of actor network and critic network at time slot $t$). In our experiment, we assume that at time slot $t$, the task received by the vehicle is first placed in the queue, and then if no task is executed, the scheduler will schedule the task according to the offloading policy.
\begin{figure}[t]
	\centerline{\includegraphics[scale=0.5]{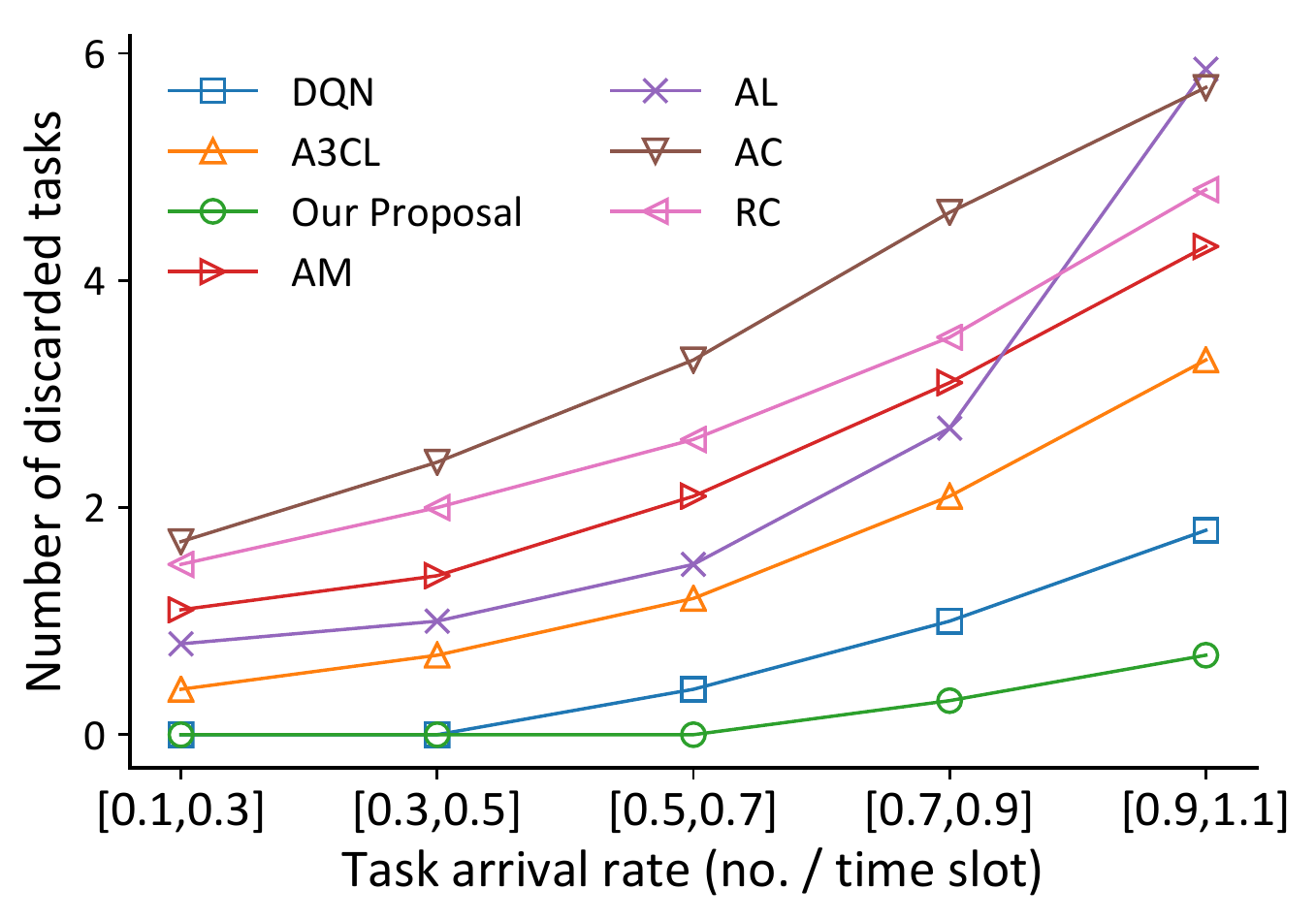}}
	\caption{Average number of discarded tasks versus task arrival rate.}
	\label{fig:diff_task_rate}
\end{figure}

We evaluate our algorithm under different computing resources required by tasks, data size of tasks, and task arrival rate. It can be seen from Fig.~\ref{fig:dynamic_required_r} that when the required computing resources is very light, AL can almost obtain the same cost as our proposal. This is because local vehicles can meet the task's demand for computing resources at this time. When the computing resources required for the task exceed the local computing power, the cost of AL is significantly increased. In addition, DQN outperforms other baselines, and our algorithm performs best. Fig.~\ref{fig:dynamic_data_size_r} compares the average cost with different data size of tasks. The cost of all algorithms except AL decreases with the increasing of data size of tasks. This is because for AL, tasks are only executed on the vehicle, and the data size of tasks has no correlation with the cost. Similar to static queue scenario, AC reaches high cost when the data size of tasks is large. In Fig.~\ref{fig:dynamic_task_r}, our proposal performs best among all algorithms. As the task rate increases, the cost of algorithms other than our proposal increases significantly, which means that the scheduler does not preserve computing resources for unexpected tasks, and a queue overflow event has occurred due to accumulation of tasks. It is worth noting that compared with A3CL, our proposal with task prediction can achieve the smallest cost, which reflects our algorithm in digital twin empowered vehicles is efficient for solving task offloading problem. 

In dynamic queue scenario, tasks may not be executed in time and cause the queue to overflow, which will lead to serious consequences. Especially when the task rate changes, the scheduler will not be able to adapt to this change according to the current state. Therefore, preserving available computing resources is particularly important in task offloading system. In addition, the speed of vehicle also affect the cost of tasks, especially at the junction of the RSU coverage area (i.e., after crossing the boundary, the vehicle will re-transmit the task, resulting in high cost). We verify the effectiveness of the algorithm under the two factors. First, we compare the performance of all algorithms under different speed of vehicle, as shown in Fig.~\ref{fig:diff_v}. It is observed that the average number of re-transmitted tasks of AC, RC, and AM grows rapidly as the vehicle speed increases. This is because for AC and AM, when the vehicle speed increases, the tasks will be interrupted during transmission due to the vehicle crossing the RSU coverage boundary. Note that the average number of re-transmitted tasks of AL remains the same (i.e., equal to 0). The reason is that all tasks are executed in the local vehicle, which will not lead to task re-transmission. However, this will cause the queue to overflow due to the accumulation of tasks in the task queue which is strengthened by Fig.~\ref{fig:diff_task_rate}. In particular, for our proposal, when the vehicle speed increases from $5$ m /s to $20$ m /s, there is no re-transmission tasks. As the vehicle speed continues to increase, the number of re-transmitted tasks of the our algorithm increases slightly, which shows that our algorithm can make the optimal offloading decision before the vehicle crosses the RSU boundary to minimize the number of interrupted tasks. 

The comparison of average number of discarded tasks under different task arrival rate is shown in Fig~\ref{fig:diff_task_rate}. As shown, AC performs the worst because offloading all tasks to the cloud server means very high transmission time, especially when the $\lambda$ increases, the queue overflow event occurs because the task cannot be scheduled in time. It is observed that the number of discarded tasks of AL increases drastically when the task rate increases from $\left[ 0.7,0.9\right]$ to $\left[ 0.9,1.1\right] $, especially when the task rate reaches $\left[ 0.9,1.1\right] $, the number of discarded tasks of AL even exceeds that of AC. The reason is that when the task rate is low, the computing resources of the local vehicle can meet the task calculation requirements. However, when the task rate is high, the limited computing power of the local vehicle cannot process a large number of tasks in time, causing the queue to overflow. A3CL performs worse than DQN and our proposal because it does not consider task prediction, i.e., the algorithm cannot reserve suitable computing resources for a large number of tasks in advance because it cannot predict the arrival of tasks, resulting in a large backlog of tasks being discarded. It can be seen that our algorithm performs best. When $\lambda$ exceeds $\left[ 0.5,0.7\right]$, a small part of tasks are discarded. This is because the task rate exceeds the workload of the system, which is not common under normal circumstances. From the figure, we can conclude that our algorithm with task prediction powered by digital twin can reserve valuable computing resources for future tasks in advance to reduce the number of discarded tasks, which will greatly improve the QoE.

\section{Conclusion}
Digital twins can provide global and historical information for vehicles in physical space through inter-twin communications. In this paper, we investigate the computing offloading scheduling problem and propose a digital twin empowered task offloading framework for IoV. Specifically, tasks can be executed on local vehicles, MEC server, or clouds. Different from previous work, we consider the change in wireless channel status during the task offloading process, and the fluctuation of the computing resources of the MEC server, which will be more in line with real conditions. In addition, we predict the arrival of future tasks through recurrent neural network to preserve computing resources and reduce overflow events caused by task accumulation. Our objective is to minimize the long-term cost in terms of a trade off between task latency, energy consumption, and renting cost of clouds. We model the optimization problem as an MDP and resort to deep reinforcement learning (DRL) to deal with it. Simulation results demonstrate that our proposed algorithm significantly outperforms the baselines. Especially with the support of digital twins, vehicles can effectively predict the arrival of tasks and reserve computing resources for this purpose.

\appendices
\section{Task Prediction module}\label{append:task_pred}
\begin{figure}[tbp]
	\centerline{\includegraphics[scale=0.47]{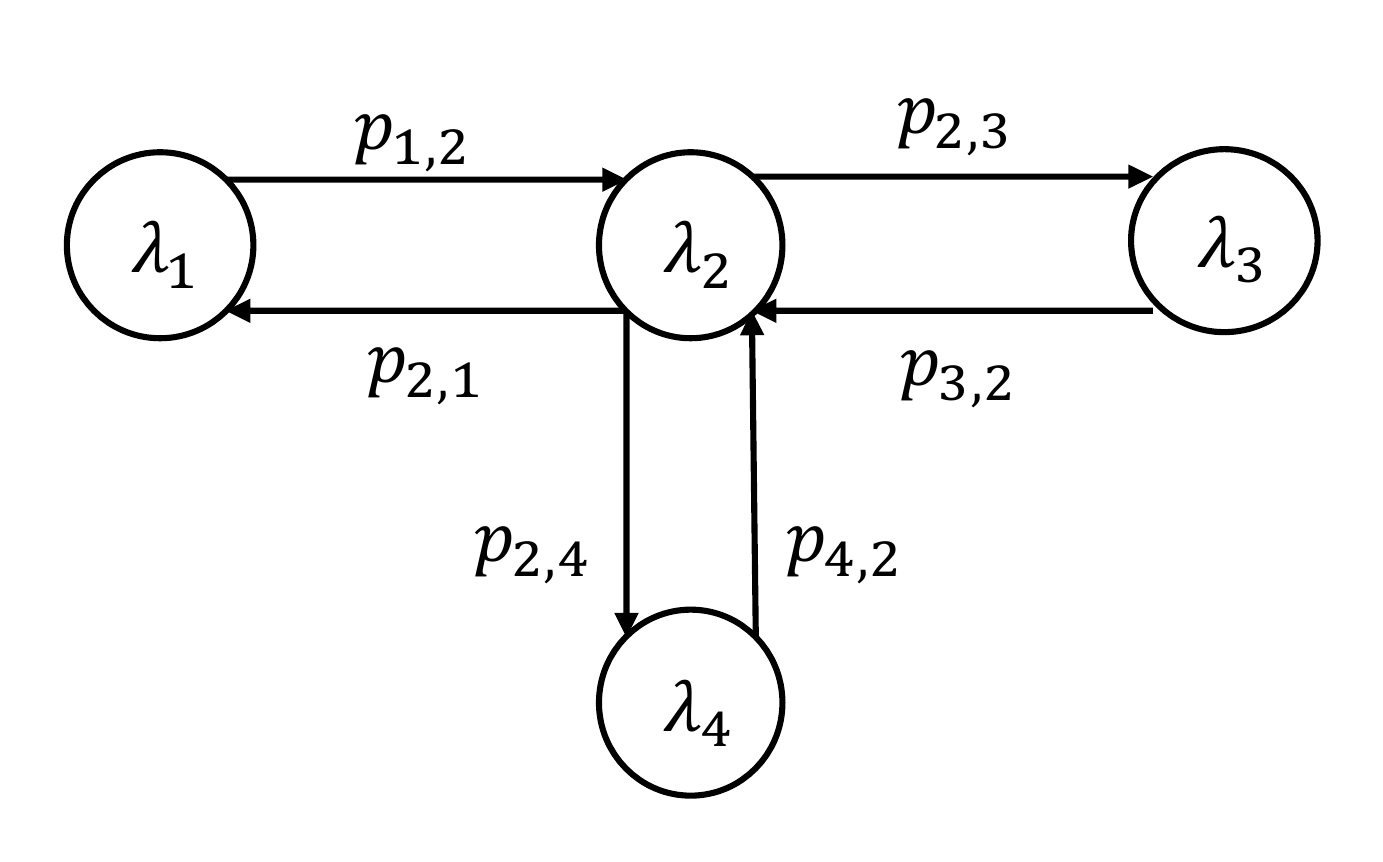}}
	\caption{An example of state space and transitions of the Markov process in model of task rate.}
	\label{fig:task_mdp}
\end{figure}

Note that the arrival rate of tasks is time-dependent (i.e., the tasks received by the vehicle are related to each other). RNN can effectively solve the time correlation between data. Therefore, the task rate prediction module is composed of a RNN unit with 128 cells and a FC layer with 512 neurons. In this module, the task rate of the past five time slots are fed into the input layer, and the output layer can predict the data rate at next time slot which is used as a part of current state to training the actor network and the critic network. The task arrival process is modeled as a Poisson process with rate $\lambda$ at time slot $t$. Generally, when a vehicle driving on a road with a complex environment, it receives a different number of tasks \footnote{At different time slots, the arrival rate of tasks may be different. For example, the vehicle receives more tasks when braking suddenly than when driving slowly.} (e.g., brake or accelerate) at different time slots. Since tasks are time-dependent, we model the arrival process of different tasks as a Markov process, and an example of the Markov process is shown in Fig.~\ref{fig:task_mdp}. In the Markov chain, the node represents the arrival rate of the task, (e.g., $\lambda_{1}$) and $p_{i,j}$ represents the transition probability from node $i$ to node $j$ following the state transition matrix.  The state transition matrix $P$ corresponding to this example is given by

\begin{equation}
	P=\left( 
	\begin{array}{cccc}
		0 & p_{1,2} & 0 & 0 \\ 
		p_{2,1} & 0 & p_{2,3} & p_{2,4} \\ 
		0 & p_{3,2} & 0 & 0 \\ 
		0 & p_{4,2} & 0 & 0%
	\end{array}%
	\right) 
\end{equation}
In our simulation, the task generation system samples a task rate (denoted as $\lambda_{i}$) from the Markov chain every $20$ time slots \footnote{We assume that in Markov chain, the arrival rate of the task  changes  every 20 time slots.} and generate tasks following $\lambda_{i}$. At the same time, these task rates will be recorded and used as input to the task prediction module. 

\section{Throughput Prediction module}\label{append:throu_pred}
In our task offloading problem, digital twins exchange data regularly to obtain global historical information, e.g., past channel gains. The digital twin of the vehicle considers these past channel gains to predict the change in throughput for a period of time in the future and adds the predicted throughput to the state vector of the MDP for DRL training. The advantage of this is that the digital twin can consider future throughput when making offloading decisions, and avoid transmitting tasks in bad wireless conditions. 

\begin{figure}[t]
	\centerline{\includegraphics[scale=0.5]{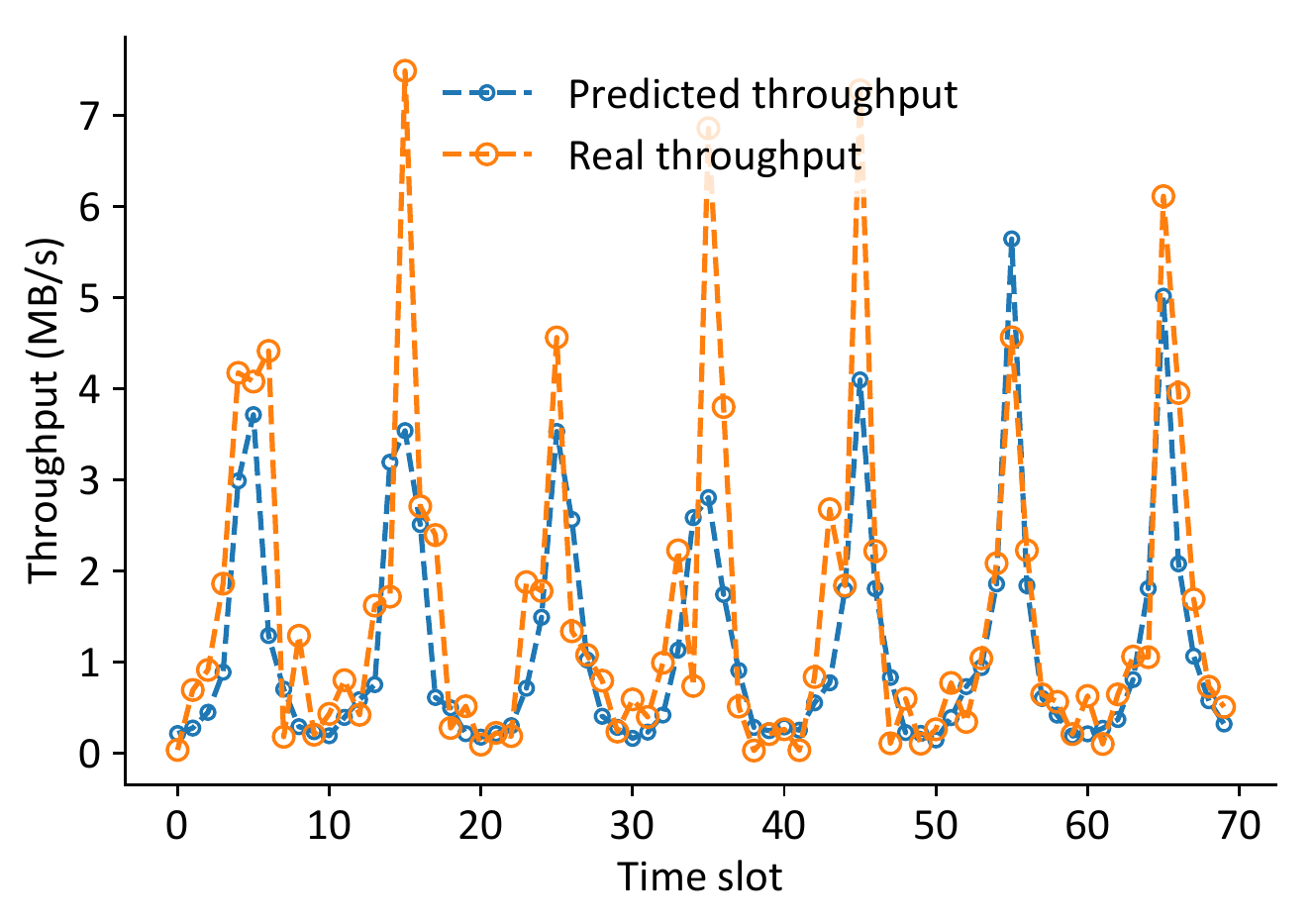}}
	\caption{Real throughput versus the predicted throughput using digital twins.}
	\label{fig:pred_throughput}
\end{figure}

The throughput prediction module is divided into throughput prediction from vehicle to cloud server and from vehicle to RSU. Each prediction module is a RNN with 256 cells and two FC layers. The input layer includes channel information of the past $50$ time slots, and the first FC layer has 512 neurons and the second FC layer has 256 neurons. Each FC layer uses sigmoid as the activation function. The output layer has 1 linear unit representing the average throughput of the predicted 10 time slots in the future. Fig.~\ref{fig:pred_throughput} shows result of the future throughput predicted by the digital twin. It is observed that the predicted result is quite in line with the true throughput with the prediction accuracy of $96.4 \%$, which demonstrates the effectiveness of the throughput prediction module.


\bibliography{IEEEabrv,mybib}{}
\bibliographystyle{IEEEtran}
\end{document}